\documentclass[sigconf]{acmart}

\AtBeginDocument{%
  \providecommand\BibTeX{{%
    \normalfont B\kern-0.5em{\scshape i\kern-0.25em b}\kern-0.8em\TeX}}}

\copyrightyear{2024}
\acmYear{2024}
\setcopyright{rightsretained}
\acmConference[CHI '24]{Proceedings of the CHI Conference on Human Factors in Computing Systems}{May 11--16, 2024}{Honolulu, HI, USA}
\acmBooktitle{Proceedings of the CHI Conference on Human Factors in Computing Systems (CHI '24), May 11--16, 2024, Honolulu, HI, USA}
\acmDOI{10.1145/3613904.3642776}
\acmISBN{979-8-4007-0330-0/24/05}

\usepackage{xspace}
\usepackage{float}
\usepackage{multirow}
\usepackage[nomessages]{fp}%
\usepackage{graphicx}
\usepackage{calc}
\usepackage{subcaption}
\usepackage{xcolor}
\usepackage{sparklines}

\begin{document}

\title[Glanceable Data Visualizations for Older Adults]{Glanceable Data Visualizations for Older Adults: Establishing Thresholds and Examining Disparities Between Age Groups}

\author{Zack While}
\email{zwhile@cs.umass.edu}
\orcid{0000-0002-9114-3984}
\affiliation{%
  \institution{University of Massachusetts Amherst}
  \city{Amherst}
  \state{Massachusetts}
  \country{USA}
  \postcode{01002}
}
\author{Tanja Blascheck}
\email{research@blascheck.eu}
\orcid{0000-0003-4002-4499}
\affiliation{%
  \institution{University of Stuttgart}
  \city{Stuttgart}
  \country{Germany}
  \postcode{}
}
\author{Yujie Gong}
\email{ygong@smith.edu}
\orcid{0009-0003-8629-3472}
\affiliation{%
  \institution{Smith College}
  \city{Northampton}
  \state{Massachusetts}
  \country{USA}
  \postcode{01063}
}
\author{Petra Isenberg}
\email{petra.isenberg@inria.fr}
\orcid{0000-0002-2948-6417}
\affiliation{%
  \institution{Universit\'e Paris-Saclay, CNRS, Inria, LISN}
  \city{Orsay}
  \country{France}
  \postcode{}
}
\author{Ali Sarvghad}
\email{asarv@cs.umass.edu}
\orcid{0000-0003-3718-7043}
\affiliation{%
  \institution{University of Massachusetts Amherst}
  \city{Amherst}
  \state{Massachusetts}
  \country{USA}
  \postcode{01002}
}

\renewcommand{\shortauthors}{While et al.}

\begin{abstract}
    We present results of a replication study on smartwatch visualizations with adults aged 65 and older. The older adult population is rising globally, coinciding with their increasing interest in using small wearable devices, such as smartwatches, to track and view data. Smartwatches, however, pose challenges to this population: fonts and visualizations are often small and meant to be seen at a glance. How concise design on smartwatches interacts with aging-related changes in perception and cognition, however, is not well understood. We replicate a study that investigated how visualization type and number of data points affect glanceable perception. We observe strong evidence of differences for participants aged 75 and older, sparking interesting questions regarding the study of visualization and older adults. We discuss first steps toward better understanding and supporting an older population of smartwatch wearers and reflect on our experiences working with this population. Supplementary materials are available at \url{https://osf.io/7x4hq/}.
\end{abstract}



\newcommand{\baricon}{\texttt{Bar}~\includegraphics[height=1.5ex]{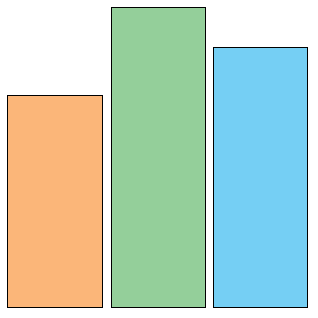}\xspace}
\newcommand{\donuticon}{\texttt{Donut}~\includegraphics[height=1.5ex]{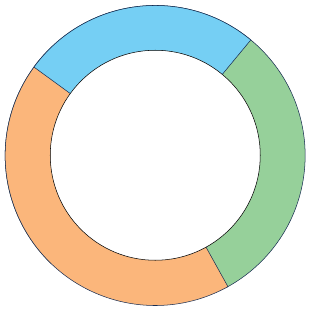}\xspace}
\newcommand{\radialicon}{\texttt{Radial}~\includegraphics[height=1.5ex]{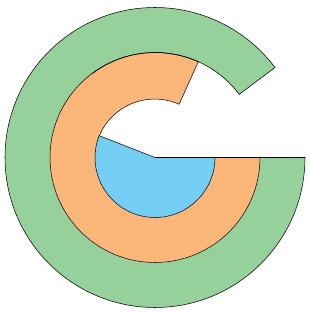}\xspace}
\newcommand{\meanicon}{\raisebox{-.8\dp\strutbox}{\includegraphics[height=2.75ex]{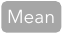}}\xspace}

\newcommand{\avgsd}[2]{(\textit{Avg.} = {#1}, \textit{SD} = {#2})\xspace}
\newcommand{\avgsdrange}[4]{(\textit{Avg.} = {#1}, \textit{SD} = {#2}, \textit{Range} = {#3}-{#4})\xspace}

\newcommand{\legendbox}[1]{%
\textcolor{#1}{\rule{\fontcharht\font`X}{\fontcharht\font`X}}%
}
\definecolor{youngeradults}{HTML}{F28E2B}
\definecolor{olderadults}{HTML}{AF7AA1}
\definecolor{youngold}{HTML}{76B7B2}
\definecolor{old}{HTML}{EDC949}
\definecolor{skipr24}{HTML}{b6252c}
\definecolor{staircaseerror}{HTML}{ff4400}
\definecolor{staircaselightblue}{HTML}{00B0F0}
\definecolor{kbfigred}{HTML}{FF0000}
\newcommand{\yadultcbox}{\legendbox{youngeradults}\xspace}
\newcommand{\oadultcbox}{\legendbox{olderadults}\xspace}
\newcommand{\yoldcbox}{\legendbox{youngold}\xspace}
\newcommand{\oldcbox}{\legendbox{old}\xspace}
\newcommand{\scerrcbox}{\legendbox{staircaseerror}\xspace}
\newcommand{\kbfigcbox}{\legendbox{kbfigred}\xspace}

\renewcommand{\sectionautorefname}{Section}
\renewcommand{\subsectionautorefname}{Section}
\renewcommand{\subsubsectionautorefname}{Section}

\definecolor{donutyoung}{HTML}{bdbdbd}
\newcommand{\maxnum}{43} 
\newlength\maxlen
\newcommand{\databar}[2][isoorange]{%
  \settowidth{\maxlen}{\maxnum}%
  \addtolength{\maxlen}{\tabcolsep}%
  \FPeval\result{round(#2/\maxnum:4)}%
  \raisebox{0pt}[0pt][0pt]{\rlap{\color{#1}\hspace*{-.7\tabcolsep}\rule[-.20\ht\strutbox]{\result\maxlen}{1.20\ht\strutbox}}}%
  \makebox[\dimexpr\maxlen-\tabcolsep][r]{#2}%
}

\definecolor{device-clr}{HTML}{66c2a5}
\definecolor{health-clr}{HTML}{fc8d62}
\definecolor{weather-clr}{HTML}{8da0cb}
\definecolor{none-clr}{HTML}{7f7f7f}
\newcommand{\devicecbox}{\legendbox{device-clr}\xspace}
\newcommand{\healthcbox}{\legendbox{health-clr}\xspace}
\newcommand{\weathercbox}{\legendbox{weather-clr}\xspace}
\newcommand{\nonecbox}{\legendbox{none-clr}\xspace}

\maketitle

\section{Introduction}
Glanceable visualizations are concise graphical representations of data primarily designed and used for enabling quick insight discovery~\textit{at a glance} without extensive exploration or analysis~\cite{pizza2016smartwatch, isenberg2021microVis,blascheck2021characterizing}. These visualizations are standard on devices with small displays, such as smartwatches and activity trackers. Most interactions with such devices are less-than-5-second peeks aimed at fast retrieval of key information~\cite{pizza2016smartwatch}. This brevity aligns well with the design goals, characteristics, and affordances of glanceable visualizations. Moreover, the constrained display size of these devices limits the utility of complex and information-dense visualizations~\cite{neshati2019challenges}, making glanceable formats a suitable fit. 

Prior research (e.\,g.,~\cite{blascheck2018glanceable,blascheck2021characterizing,blascheck2021replication,blascheck2023studies}) has investigated various aspects of glanceable visualizations' design and utility. However, a crucial gap remains in our understanding of how these findings apply to \textit{older adults}.\footnote{In line with the World Health Organization, we use the term ``older adults'' to refer to individuals aged 65 years and over.}  This knowledge gap becomes even more critical in light of the fast-growing population of older adults and the rapid emergence of visualization tools and technologies to assist older adults with data-driven self-care and decision-making~\cite{backonja2016visualization}.

As individuals age, they undergo gradual changes in perception, cognition, and physical abilities that can impact their capacity to use visualizations effectively. For example, a recent survey by Fan et al.~\cite{fan2023understanding} found that older adults struggled to effectively use and gain insights from COVID-19 visualizations due to factors such as indistinguishable colors for aging vision (e.\,g., blue-green), low contrast ratios between graphical elements, and small font sizes. The onset and progression of age-related changes can vary greatly among individuals due to genetics, lifestyle, and overall health. Nevertheless, the physiological process of aging affects everyone universally. For instance, age-related farsightedness (presbyopia) is nearly ubiquitous in adults 65 years and older~\cite{johnson2017designing}, and many people experience a decline in visual acuity above age 50~\cite{mitzner2015considering}. Certain cognitive functions such as memory, attention, and processing speed may also decline with age, which can influence information processing and decision-making~\cite{unsworth2007nature}. The compounded effects of aging may affect older adults' use of glanceable visualizations on small screens and, therefore, the usability of smartwatches. However, the intersection of glanceable visualization and aging is a notably under-explored area of research. Our understanding of the performance, preferences, and requirements of older adults, as well as the factors that may influence and shape them, remains limited.

This work seeks to narrow our current knowledge gap at the intersection of aging and visualization. We, therefore, conducted a replication of the perceptual study by Blascheck et al.~\cite{blascheck2018glanceable}, in which the authors examined and established \textit{time thresholds} across the combination of three glanceable visualization designs (\baricon, \donuticon, and \radialicon) and three data sizes (7, 12, and 24 data points) on a smartwatch. In perceptual studies, the term ``threshold'' commonly refers to a performance boundary (or limit), such as the minimal time needed for individuals to detect or distinguish information within a visualization. The authors considered a single comparison task, in which participants were required to identify two target elements (e.\,g., two bars) in the visualization and select the one representing a larger value. In our replication study, we followed the same design and procedure outlined in the original work, \textit{except for the age group of the participants}. While Blascheck et al.'s participants were predominantly younger individuals (19-64 years old), we specifically recruited participants from an older age group (65 and older) for our study. Our work primarily aimed to (1)~examine the potential variations and similarities in performance and preferences between older adults and younger individuals, (2)~investigate the extent to which aging affects performance and preference within the ``older adult'' group, and (3)~establish empirical thresholds for the speed at which older adults can execute a simple data comparison task that involves reading a glanceable visualization.

We contrast the results of our study with those of Blascheck et al., allowing for a comprehensive understanding of the distinctions between the two study age groups. In our threshold assessments, older adults exhibited the fastest overall performance with \donuticon (312 ms), followed by \baricon (485 ms) and \radialicon (2211 ms), a ranking that aligns with Blascheck et al.'s study with younger participants. Despite this similarity, older adults were consistently slower across all nine experimental conditions. Further analysis of confidence intervals' overlap showed strong (5/9 conditions) to weak (2/9) evidence of differences between younger and older adults. The performance gap widened with increasing data size (7$\rightarrow$12$\rightarrow$24) for all visualization types, hinting at a steeper performance decline in older adults as visual complexity grew. To gain a more nuanced understanding of the relationship between the progression of age and glanceable visualization performance, we compared the performance of the younger adults from the prior study by Blascheck et al. ($n=18$) with two older age segments: ``young-old'' (age 65-74, $n=12$) and ``old-old'' (age $\geq 75$, $n=12$).\footnote{Terms young-old and old-old are used in gerontology to distinguish between different segments of the older population (e.\,g.,~Baltes and Smith~\cite{baltes2003new}).} We followed conventional gerontological practice by setting the cut-off age for the old-old category at 75 (e.\,g.,~\cite{abdel1997consumption, sinoff1997barthel,baltes2003new}).

We observed minimal performance differences between younger adults and the young-old in 8 out of 9 conditions. However, the performance discrepancies increased with age, with evidence of several differences between the old-old and the two other age groups. These findings suggest that the impact of aging on visualization performance may accelerate and intensify with advanced age.

This work makes several contributions to the visualization field. First, we fill a gap by quantitatively establishing time thresholds for older adults in the context of glanceable visualizations, enriching the current understanding of how age impacts visual information processing. Second, we offer evidence that underscores the influence of age on graphical perception. Our findings demonstrate that \textit{age matters} and should be incorporated into both research and practices in visualization. Lastly, reflecting on the lessons learned conducting this work, we outline considerations for designing and conducting human-centered studies with older adults. These contributions collectively set the stage for further investigations into the complex interplay between aging and visualization.
\section{Motivation for Focusing on Older Adults and Replication}
This section addresses three key considerations: (1) The rationale for focusing on older adults, (2) the reasons for conducting a replication study, and (3) the choice of Blascheck et al.'s~\cite{blascheck2018glanceable} study.

\subsection{Why Older Adults?}
The knowledge gap at the intersection of data visualization with aging is significant and under-researched~\cite{backonja2016visualization,le2016, brandt2014replication}. This issue is gaining urgency in light of the growing global population of older adults, which is expected to rise to 16\% by 2050 and 24\% by 2100; in the U.S., older adults will outnumber children by 2060~\cite{Demograp46:online}. The use of visualizations in enhancing older adults' lives, such as in health monitoring wearables, is on the rise, with most interventions including visualizations~\cite{cajamarca2020}. However, challenges in design and usability hinder adoption among older adults~\cite{chung2023community}. Addressing this research area is timely and ethically necessary, responding to calls for more diverse, inclusive, and equitable research in visualization~ \cite{lee2020reaching, angerbauer2022toward, marriott2021inclusive}.

\subsection{Why Replication?}
Our research aligns with a \textit{conceptual replication}~\cite{brandt2014replication, crandall2016scientific, stroebe2014alleged}, a replication study type, in which the same research questions or hypotheses are examined as in the original study but with \textit{controlled changes} in methods, participants, or settings. Conceptual replication aims to test and extend the generalizability of prior findings. Crandall and Sherman~\cite{crandall2016scientific} argue for its importance in scientific research by emphasizing its role in testing theories across various settings and conditions, contributing to a broader understanding of the subject. Unlike direct replications, which primarily seek to confirm original findings, conceptual replications challenge and expand these theories. They are more efficient in advancing the field and identifying theories' limitations or boundary conditions. The ongoing debate in the scientific community about direct versus conceptual replications is noted; for example, Stroebe and Strack~\cite{stroebe2014alleged} suggest that conceptual replications, by testing a theory's validity across different settings and populations, offer more robust evidence than direct replications. Throughout this paper, \emph{replication} specifically refers to \emph{conceptual replication}.

\subsection{Why Blascheck et al.'s Study?}
Previous studies on smartwatch perception mainly involved young participants (e.\,g.,~\cite{blascheck2018glanceable,blascheck2023studies, Neshati2019}). Given the smaller displays of smartwatches, age-related vision changes, and the growing interest of older adults in smartwatches, it is crucial to determine if these findings apply to them. Our replication of Blascheck et al.'s study~\cite{blascheck2018glanceable} was motivated by four factors: its focus on fundamental aspects of smartwatch perception, the inclusion of basic comparison tasks relevant to many visualization tasks, the use of various visualization types and data sizes, and the prior study's participant age range ($<65$), enabling direct performance comparisons between older and younger adults. Additionally, its previous replication with young participants on larger displays confirmed the original results' robustness~\cite{blascheck2021replication}. Lastly, access to detailed information from the original researchers facilitated a comprehensive replication.
\section{Related Work}

While prior work on glanceable visualization and aging is remarkably sparse~\cite{Cajamarca2023understanding}, the existing literature provides a starting point that informs and motivates this study. In this section, we discuss relevant current work on glanceable visualization (\autoref{sec:glance-vis}), visualizations for older adults (\autoref{sec:vis-oa}), as well as smartwatches and older adults (\autoref{sec:sw-oa}).

\subsection{Glanceable Visualizations}
\label{sec:glance-vis}
Glanceable visualizations depict information that can be gleaned \emph{at a glance}, focusing on design choices that communicate data as concisely as possible, although different areas of research present varying definitions of the length of a ``glance''~\cite{blascheck2021characterizing}. Consolvo et al.~\cite{consolvo2008flowers} presented one of the earliest works on glanceable displays, demonstrating the positive impacts of a mobile fitness tracker on wearers' motivation. Blascheck et al.~\cite{blascheck2023studies} observed that watch faces with a digital time representation and multiple proportion visualizations lead to better performance with bar and radial bar charts and outperformed text depicting progress toward a target value. Bar charts led to higher accuracy. However, participants preferred radial bar charts concerning aesthetics. Adding distracting elements, such as an analog time representation, did not significantly affect performance. Horak et al.~\cite{horak2018david} investigated the interplay of smartwatches and large interactive displays in data analysis scenarios, using smartwatches to store and show sets of data points or visualization configuration options as well as to control aspects of the large display. However, that work viewed the smartwatch as a supplement to large displays instead of creating visualizations for smaller screens. Islam et al.~\cite{islam2022preferences} examined the design of sleep visualizations on smartwatches and fitness bands, observing greater preference for visualizations over text depictions and similar chart preferences across both devices.

Neshati et al.~\cite{Neshati2019} incorporated glanceable sparkline designs on smartwatches.
Condensed line graphs along the x-axis improved accuracy in judging line heights, reducing the number of flick operations needed to view entire graphs and maximizing screen real estate. Similar efforts in efficient use of screen space include the simplified designs of Space-Filling Line Graphs~\cite{Neshati2021} as well as the work of Chen~\cite{chen2017visualizing}, which proposed utilizing smartwatch borders for interactive broad views of large time-series data. Our study focuses on older adults as a target population, while these targeted a more general audience and included significantly younger participants.

Understanding the data viewed on smartwatches is crucial, with previous studies exploring common and health-related preferences \cite{islam2020visualizing,chung2023community}. Islam et al.~\cite{islam2020visualizing} found health and fitness data as the most frequently displayed, with an average of five items on participants' watch faces at any time. Chung et al.~\cite{chung2023community} noted strong interest in step count, walking distance, and heart rate but lower interest in oxygen levels, stairs climbed, and sleep. Glanceable visualization has been studied in application-focused contexts such as qualitative exercise feedback~\cite{Gouveia2016,Schiewe2020}, while other works observed preferences for visualizing data rather than presenting it as a notification~\cite{Schiewe2020} or a number~\cite{amini2017data}.
In contrast to this past work, our perceptual study is data- and application-agnostic.

\subsection{Visualization and Older Adults}
\label{sec:vis-oa}
Research on visualization often overlooks older adults, a gap highlighted by Cajamarca et al.~\cite{cajamarca2020}. One visualization area considered for older adults is health data monitoring and decision-making. This field explores the impact of visualization systems on life decisions, particularly for older adults. Galesic et al.~\cite{Galesic2009} found that icon arrays aided risk comprehension in individuals with low numeracy, including older adults. Pham et al.~\cite{Pham2012} observed that health visualizations assisted older adults in progressing toward health goals, while Vargemidis et al.~\cite{vargemidis2023performance} reported increased motivation for physical activity in older adults through visualizations that emphasized enjoyment. Additionally, Price et al.~\cite{Price2016} explored how visualization affects working memory and decision-making in older adults. They found that color-based visualizations were more effective than tabular data for medical decision-making, although this advantage decreased in more complex situations. Collectively, these studies underscore the potential of visualization in enhancing the decision-making and well-being of older adults, including applications with smartwatches.

Other work has focused on overall \emph{visualization design} with older adults in mind. Le et al.~\cite{Le2012} evaluated health visualization designs, including stacked bar charts, wellness polygons, and partitioned donut charts. They found that while bar charts were familiar, they demanded high cognitive load; polygons were complex for trend analysis; and partitioned donut charts, though adequate for trend depiction, also required high mental effort. Subsequent studies (Le et al.~\cite{le2016, Le2015elementary}) showed older adults' preference for separate line graphs over stacked ones, and a longer time was taken for value comparison than younger adults. Le et al.~\cite{le2015evaluation} further discovered that multiple visual cues in health visualizations could impair older adults' information processing, with participants using a combination of prior knowledge and visualization elements when interacting with visual displays. Fan et al.~\cite{fan2023understanding} identified challenges for older adults in understanding interactive COVID-19 data visualizations, such as insufficient information clarity. Additionally, Cajamarca et al.~\cite{Cajamarca2023understanding} observed that older Chilean adults interpreted health data on smartwatches more accurately without progress indicators, which otherwise distracted them from essential information. They also found that older adults with higher technology proficiency had better accuracy and speed in data interpretation. These studies highlight how aging affects visualization perception, a challenge that may be compounded when dealing with small-scale smartwatch displays.

The existing but limited body of work on visualization for older adults highlights critical aspects of design and graphical perception specific to this demographic. However, only one previous work was performed using smartwatches or glanceable visualizations~\cite{Cajamarca2023understanding}. This gap in the current literature underscores the pressing need for further exploration to determine whether the established understanding of graphical perception remains applicable in the face of significant shifts in screen size (small) and design objectives (rapid information retrieval).

\subsection{Smartwatches and Older Adults}
\label{sec:sw-oa}
Rosales et al.~\cite{rosales2017older} assessed the evolving attitudes and behaviors of older adults using smartwatches over time. Participants with a preexisting interest in technology showed greater enthusiasm for learning new technology, while others envisioned the smartwatch potentially replacing some smartphone functions. Li et al.~\cite{li2020factors} found that prolonged wearable device usage positively impacted older adults' health.  Consistent daily usage formed long-term habits, corresponding with improved health outcomes. These findings highlight the potential of smartwatches to enhance older adults' lives, encouraging behaviors that boost their well-being and cultivate interest in other technologies.

Khakurel et al.~\cite{Khakurel2018} investigated smartwatch usability issues for older adults, identifying persistent challenges such as screen size, typography, tap detection, and interaction techniques. Notably, screen size and font size emerged as critical factors influencing smartwatch usage. Chung et al.~\cite{chung2023community} found positive attitudes among older adults toward wearable devices for activity tracking but highlighted potential barriers like screen design, size, and complexity. Cristescu et al.~\cite{Cristescu2022} proposed a model exploring factors shaping older adults' behavioral intentions towards wearable technology, observing that design aesthetics, performance expectancy, effort expectancy, and facilitating conditions played significant roles. This underscores the impact of external support on older adults' behavioral intentions. Collectively, these works emphasize the need for nuanced design considerations to enhance smartwatch usability and appeal for older wearers, motivating our replication study as a foundational step toward more usable smartwatch visualizations.
\section{Study Design and Execution}
\label{sec:study-design-exec}
This study is a conceptual replication of the \textit{Evaluating Random Differences} study by Blascheck et al.~\cite{blascheck2018glanceable} detailed in Section 6 of the original paper, with the notable difference that all participants in the current study were aged 65 or older. Unless otherwise stated, all details regarding the study design and execution were replicated according to the original research to the best of our ability.

\subsection{Participants}
After preregistration, approval of the study by our institution's IRB, and requesting permission from local older adult community groups, we posted a PDF advertisement flyer to several email lists and a few in-person advertisement boards. We recruited 24 participants (19 female, 4 male, and one preferred not to answer), with ages ranging from 65 to 96 years old~\avgsd{73}{8}. Participants' education levels were high school diploma or less, associate's degree, bachelor's degree, master's degree, and doctorate, with almost all (22) participants possessing at least a bachelor's degree. All participants had normal or corrected-to-normal vision (self-reported), only one participant did not wear glasses, and no participant had a color vision deficiency (self-reported).

Among the 24 participants, 13 reported having no prior familiarity with visualizations, and their years of experience ranged from 0 to 70~\avgsd{18}{26}. 
Participants were asked for their familiarity with each chart on a Likert scale~\cite{likert1932technique} from 1 (not at all) to 5 (very familiar), reporting on \baricon~\avgsd{4.1}{1.1}, \donuticon ~\avgsd{3.2}{1.6}, and \radialicon ~\avgsd{1.5}{1.0}. 
New for this study to investigate its possible effect on performance, participants were asked to rate their tech-savviness on a Likert scale from 1~(strongly disagree to consider myself a tech-savvy person) to 5~(strongly agree)
~\avgsd{2.6}{1.1}. 
All participants finished the study, which typically took between 45 to 75 minutes. Participants then received a \$25 Amazon gift card. Comparisons of this study's participant demographic information to those of the original research are provided in \autoref{tbl:pt2}.

\begin{table}
    \centering
    \caption{(a) Comparison of age, visualization experience (in years), and familiarity (1-5) with \baricon, \donuticon, and \radialicon charts between younger (age $< 65$) and older adults (age $\geq 65$). (b) Comparison of education levels between younger and older adults.}
    \vspace{-1em}
    \label{tbl:pt2}
    \begin{subtable}[b]{\columnwidth}
        \centering
        \resizebox{\textwidth}{!}{%
        \begin{tabular}{ccccccccccc}
        \hline
        \textbf{} & \multicolumn{2}{c}{\textbf{Age}} & \multicolumn{2}{c}{\textbf{Vis Exp.}} & \multicolumn{2}{c}{\textbf{\baricon Famil.}} & \multicolumn{2}{c}{\textbf{\donuticon Famil.}} & \multicolumn{2}{c}{\textbf{\radialicon Famil.}} \\
        \textbf{Group} & \textbf{Mean} & \textbf{SD} & \textbf{Mean} & \textbf{SD} & \textbf{Mean} & \textbf{SD} & \textbf{Mean} & \textbf{SD} & \textbf{Mean} & \textbf{SD} \\ \hline
        \multicolumn{1}{r}{Younger} & 35.0 & 13.0 & 4.5 & 1.6 & 4.6 & 1.0 & 4.3 & 1.1 & 2.3 & 1.7 \\
        \multicolumn{1}{r}{Older} & 73.3 & 7.8 & 17.9 & 26.1 & 4.3 & 1.1 & 3.3 & 1.6 & 1.6 & 1.0 \\ \hline
        \end{tabular}%
        }
        \caption{}
        \label{tbl:demog-comp}
    \end{subtable}
    \hfill
    \begin{subtable}[b]{\columnwidth}
        \centering
        \resizebox{\textwidth}{!}{%
        \begin{tabular}{rccccc}
        \hline
        \multicolumn{1}{c}{\textbf{Group}} & \textbf{High School} & \textbf{Associate's} & \textbf{Bachelor's} & \textbf{Master's} & \textbf{PhD} \\ \hline
        Younger&18\% & 0\% & 12\% & 71\% & 0\% \\
        Older & 4\% & 4\% & 29\% & 38\% & 25\% \\ \hline
        \end{tabular}%
        }
        \caption{}
        \label{tbl:educ-lev-comp}
    \end{subtable}
\end{table}

\subsection{Apparatus}
\begin{figure*}
    \centering
    \begin{subfigure}[b]{0.32\textwidth}
        \centering
        \includegraphics[width=0.9\textwidth]{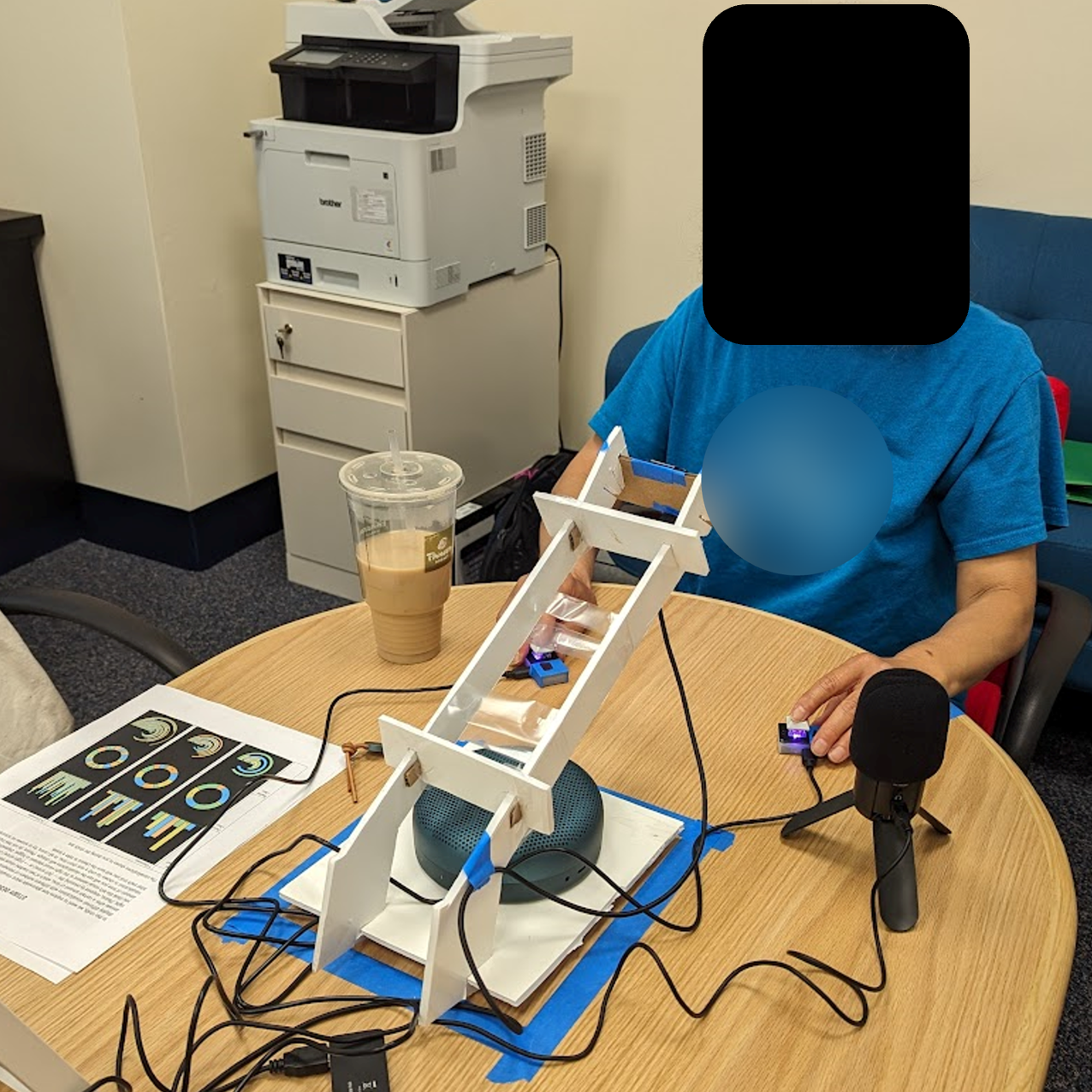}
        \caption{}
        \Description{A picture of the study setup. A participant sits at a table with the watch stand, fingers ready to press the two separate buttons (left and right). A microphone sits nearby, facing the participant.}
        \label{fig:particip-ex}
    \end{subfigure}
    \hfill
    \begin{subfigure}[b]{0.32\textwidth}
        \centering
        \includegraphics[width=0.9\textwidth]{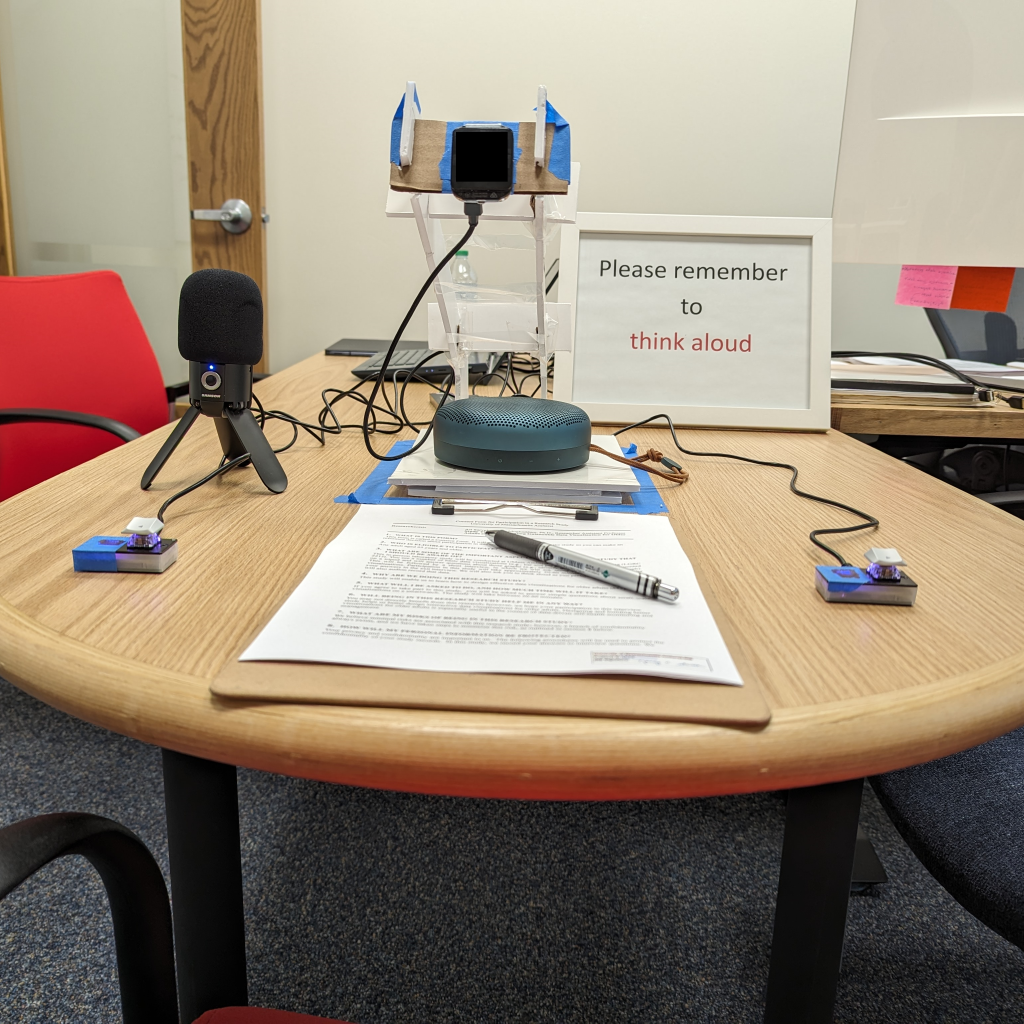}
        \caption{}
        \Description{An image of the study setup from the participants perspective. The pre-study questionnaire sits on a clipboard, and the watch is visible on the watch stand facing the participant (camera). A sign saying "Please remember to think aloud" is visible behind the stand.}
        \label{fig:particip-pov}
    \end{subfigure}  
    \hfill
    \begin{subfigure}[b]{0.32\textwidth}
        \centering
        \includegraphics[width=0.9\textwidth]{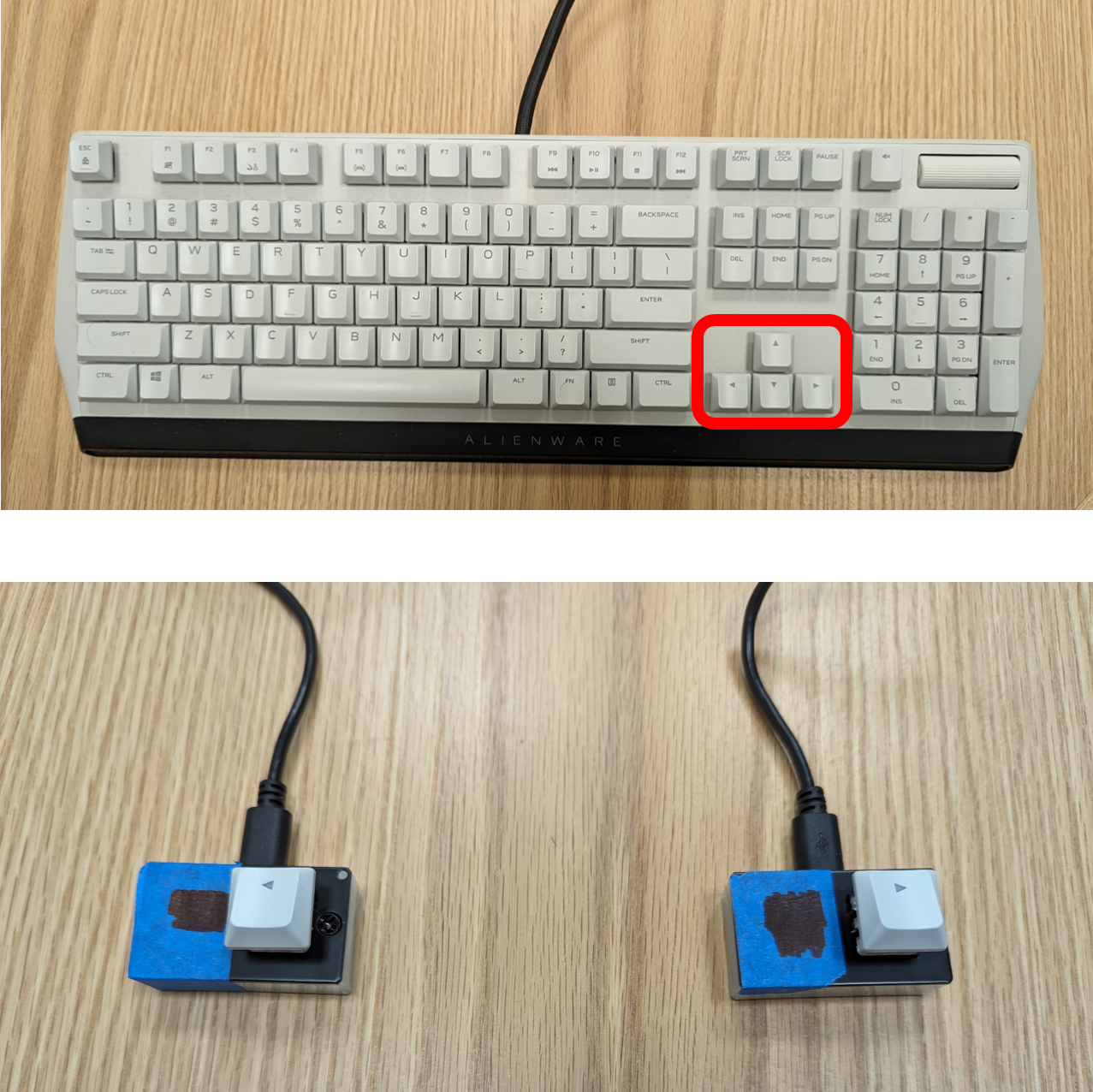}
        \caption{}
        \Description{On the top portion, the mechanical keyboard used in the study is shown. A red box is around the four arrow keys in the standard of the bottom-right portion of the keyboard. On the bottom are two separate buttons, each with its USB cable. They look like individual keys one would have on a mechanical keyboard. The keycap on the left button has the left arrow symbol printed on it, and the right button’s keycap has the right arrow printed on it.}
        \label{fig:user-periph}
    \end{subfigure} 
    \vspace{-1em}
    \caption{Study setup and apparatus: (a)~An example participant during the study. (b)~A participant's point of view. (c)~The keyboard used at the start of the study, arrow keys emphasized with a red~\kbfigcbox outline (top) and the two separate buttons (bottom) added after P6.}\label{fig:particip_ex_and_pov}    
\end{figure*}

We used a Sony SmartWatch 3 to enable comparison with the original study. 
The watch's screen resolution is $320 \times320$ pixels, measuring 1.6 inches on each side~\cite{sonysmartwatch}. We reconstructed the watch stand (shown in \autoref{fig:particip_ex_and_pov}) using the original authors' designs, approximating a typical viewing angle. As per the original study design~\cite{blascheck2018glanceable}, the stand was placed 28 cm horizontally from the end of the table and 20 cm vertically from the table's surface. After observing some initial difficulty with seeing the watch in the pilot studies, we allowed participants to horizontally move the stand as needed~\avgsdrange{25.2}{4.9}{10.2}{28} but kept the vertical height consistent.

\subsection{Task and Stimuli}\label{sec:task-and-stimuli}
This study followed a two-alternative forced choice (2AFC) approach~\cite{macmillan2004detection}, in which participants were presented with a visualization (stimulus) comprised of one of three possible chart types (\baricon, \donuticon, or \radialicon) showing three possible data sizes (7, 12, or 24). On the stimulus, two black dots indicated two specific elements (e.\,g., two bars of a bar chart). After the visualization was shown, we asked participants to choose which dot (\textit{left} or \textit{right}) marked the larger-valued element. This meant the taller bar in \baricon, the larger region in \donuticon, or the more complete circle in \radialicon. The original study's authors provided us with the same software and set of stimuli as in the original study~\cite{blascheck2018glanceable}. A total of 396 images were provided, with half~(198) having a larger value on the left and half having a larger value on the right. 

\begin{figure*}
    \centering
    \includegraphics[width=\textwidth]{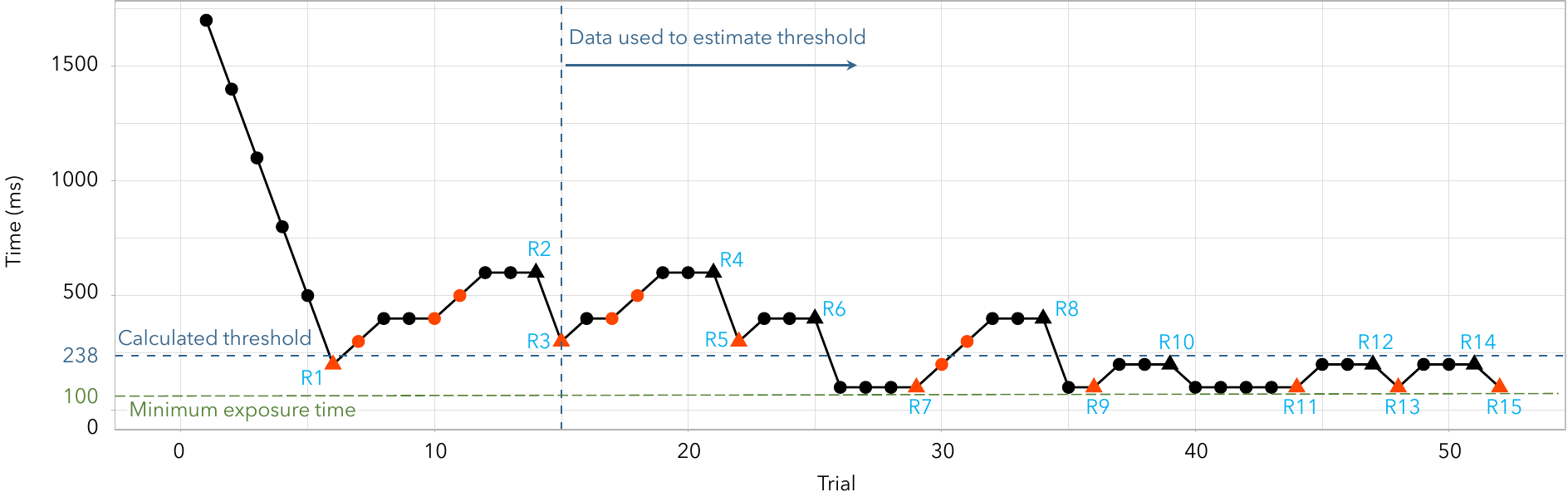}
    \vspace{-2em}
    \caption{An example of the staircase method, showing P24 performance in the \donuticon~12 condition. Each circle or triangle glyph is a trial the participant performed. The color orange-red \scerrcbox marks errors while triangle shapes mark \textit{reversals}, i.\,e., trials when a participant switched from \textit{up} (most recently increased exposure time) on the staircase to \textit{down} (decreasing exposure time), or vice-versa. Reversal points \textcolor{staircaselightblue}{R1-R15} are labeled. To compute the time threshold, the time-per-stimulus for the reversal points is averaged, starting with the third reversal~(\textcolor{staircaselightblue}{R3}). The computed threshold in this example is 238 milliseconds. The minimum exposure time was 100 milliseconds.}
    \Description{Line chart representing the staircase method for one participant, with exposure time (ms) on the Y axis and trial number on the X axis. Time ranges from 0 to 2000~ms, while trial number ranges from 0 to 55. The first trial starts at nearly 2000~ms, dropping to around 238~ms for the first reversal. After that, the line fluctuates from increasing until the subsequent reversal, then decreasing, oscillating around a line Y = 238, which is the calculated threshold.}
    \label{fig:staircase-example}
\end{figure*}

As per the original study, the stimulus display time was adjusted in real-time as a function of the participant's response using a weighted 3-down, 1-up staircase procedure that increased exposure time by 100~ms~($\Delta^{+}$)  after an incorrect response and decreased it by 300~ms~($\Delta^{-}$) after three consecutive correct responses. This results in an expected accuracy of approximately $63\%$~\cite{garcia1998forced}, though it is of note that the original study incorrectly calculated this value as approximately $91\%$~\cite{blascheck2018glanceable}. For each of the nine conditions (3 chart types $\times$ 3 data sizes), participants completed a set of trials, which we refer to as a \textit{staircase}. \autoref{fig:staircase-example} shows an example staircase from our study. A staircase would continue until the 15\textsuperscript{th} reversal (i.\,e., change from increasing to decreasing exposure time or vice-versa) or the end of the stimuli (the 198\textsuperscript{th} trial) was reached, as is typical for staircase procedures~\cite{garcia1998forced}.

\subsection{Pilot Studies}
Before the main study, we conducted four pilot studies involving volunteer older adults (ages 61-69). These pilots aimed to assess the study design and procedure, which were not tailored to older adults. During the pilots, we examined the visibility of dot markers on stimuli, the study duration, potential fatigue, eye strain, comprehensibility of the required comparison task and visualizations, and the placement and distance of the smartwatch stand.

We observed some pilot study participants encountering difficulties with \radialicon, mistakenly assuming they were measuring \textit{bar length} rather than \textit{circle completeness}. To address this, we included additional information and examples on interpreting these charts in the pre-study explanation of stimuli and tasks.

\subsection{Study Procedure}\label{study_procedure}

\begin{figure*}
    \centering
    \includegraphics[width=\textwidth]{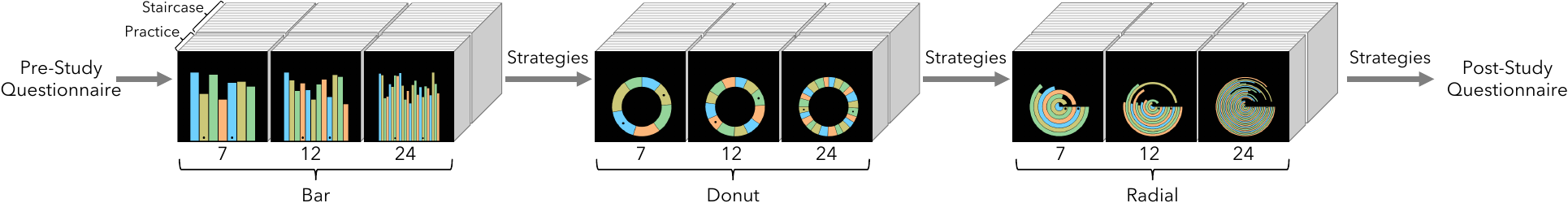}
    \vspace{-2em}
    \caption{Overall structure of the study. Each participant started and ended their study with a questionnaire. In this example, each chart type presents 7 data items for a staircase, then 12, and then 24; for each participant, this ordering was based on the Latin square design. Each staircase comprises 10 practice trials, and then the staircase, with the number of trials in the staircase depending on performance. After each chart type, we asked participants to describe strategies they used for that set.}
    \Description{Flowchart showing the set of steps in the study for a single participant. First is the pre-study questionnaire, then practice trials and staircases for the Bar 7 condition, Bar 12 condition, and then Bar 24 condition; this ordering of data sizes is just an example here, with the actual ordering chosen by the Latin Square design. This set of three staircases and a strategy question are repeated for the Donut and Radial charts. The last step is the post-study questionnaire. Example study stimuli for each condition (chart and data size combination) are also depicted.}
    \label{fig:study_design_and_procedure}
\end{figure*}

The overall study procedure is detailed in \autoref{fig:study_design_and_procedure}. Participants would fill out the pre-study questionnaire upon confirming their consent to participate. Then, an investigator would explain the study format, including how to read and compare values on each chart using a printed set of examples. 
We also told participants they could take short breaks anytime, especially between conditions.

Participants were assigned a participant ID number based on the order of appointments, which would inform the specific ordering of the chart types and data sizes; orderings were counterbalanced by a reduced Latin square design~\cite{bradley1958complete}. Because our study ($n=24$) had more total participants than the original study's Latin square design accounted for ($n=18$), there were some repeats in type-size orderings. To best allow comparison between the subgroups of \textit{young-old} participants (age $65\text{-}74$, $n=12$) and \textit{old-old} participants (age $\geq 75$, $n=12$), we made sure that there were no repeats of type-size orderings within either subgroup. 

\begin{figure*}
    \centering
    \includegraphics[width=0.75\textwidth]{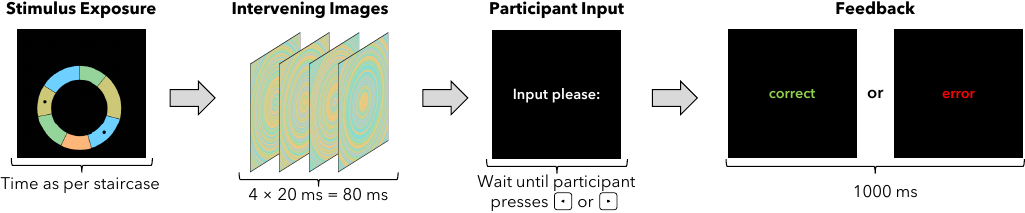}
    \vspace{-1em}
    \caption{Steps involved in an individual trial. A participant is first shown the stimulus (a chart) for an amount of time, which the current staircase determined. Then, they are shown four intervening images for 20~ms each. Next, the screen prompts the participant to either press the left or right arrow key, after which the participant is told whether their input was correct or not.}
    \label{fig:trial-details}
    \Description{Flowchart depicting the set of events in an individual trial. First, the user is shown a stimulus (chart) for an amount of time depending on the staircase, then four intervening images (total of 80~ms), then the software waits for the user to press the left or right arrow, and then the screen showing lets the user know if they were correct or not, displaying the word “correct” or “error”.}
\end{figure*}

A condition would begin with ten practice trials, with the initial exposure time between 1700~ms and 5100~ms depending on the condition; for comparability, we use the exact starting times as the previous study~\cite{blascheck2018glanceable}. The individual trial procedure is shown in \autoref{fig:trial-details}. A trial would begin with viewing the visualization, after which a set of four intervening images would show on the screen for 20~ms each to prevent afterimages of the stimulus. Then, the screen would prompt the participant to give their input (\textit{left} or \textit{right}), and lastly, the participant would be shown if their answer was correct or not for 1000~ms. After completing three staircases for one chart type, we would ask the participant what strategies (if any) they used while completing the tasks for these charts. 

\subsubsection*{Study Design Modifications}\label{modifications}
\begin{figure*}
    \centering
    \includegraphics[width=0.98\textwidth]{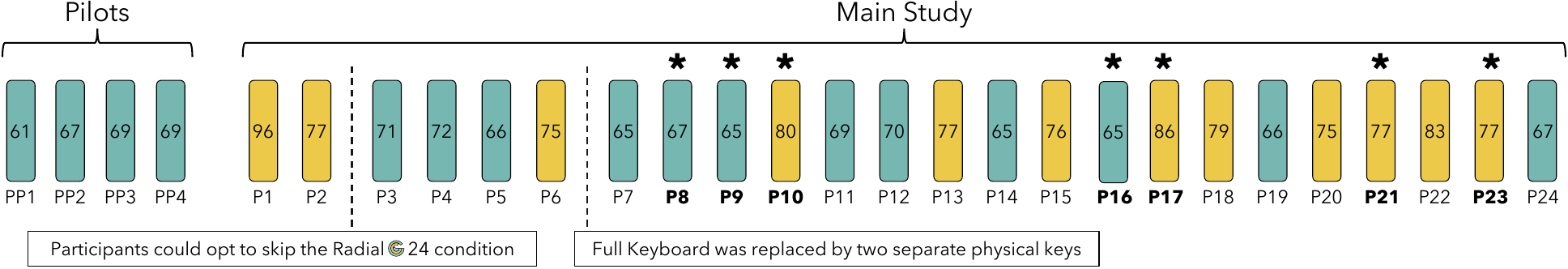}
    \vspace{-1em}
    \caption{High-level depiction of the study process and progression. We started with four pilot studies and proceeded to the main study with 24 participants. Each box in the graph represents a participant and their age. Colors indicate if that participant was in the young-old~\yoldcbox (age 65-74) or old-old~\oldcbox (age $\geq 75$) group. After the second study (P2), participants were explicitly told at the start that they could skip the \radialicon~24 condition if they were experiencing excessive fatigue, discomfort, or difficulty. Seven of the 22 remaining participants (marked with a \includegraphics[width=1.5ex]{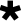}) chose to skip that condition. 
    After the sixth study (P6), we swapped the keyboard with two separate physical keys for greater ergonomic comfort during the study (shown in \autoref{fig:user-periph}).}
    \Description{Visualization of the overall study progression regarding users and their ages. The four pilot studies are first (all young-old adults), and then the 24 participants of the main study are shown (a mix of young-old and old-old participants). After the second participant (P2), participants were told they could skip the Radial 24 condition. After the sixth participant (P6), the full keyboard was replaced by two separate physical keys. Participants P8, P9, P10, P16, P17, P21, and P23 chose to skip the Radial 24 condition.}
    \label{fig:particip-skip}
\end{figure*}

 During the studies, we made two minor adjustments to the design. The first change occurred after the second study when we noticed that participants (P1 and P2) expressed concern with the difficulty of performing the \radialicon~24 condition. As a result, we decided to permit subsequent participants to skip this condition if they found it infeasible. 
Seven of the 22 remaining participants quit the \radialicon 24 condition after attempting a few trials (refer to \autoref{fig:particip-skip} for more details). Additionally, during the initial sessions of the main study, some participants expressed discomfort and difficulty with pressing the arrow keys on the full-sized keyboard used for recording their responses. To address this issue, we replaced the keyboard with two separate mechanical keys—one on the left and one on the right (refer to \autoref{fig:user-periph}). This modification seemed to improve the ergonomic experience for the participants without changing the study format. 

Notably, none of these issues emerged during our pilots, highlighting the known heterogeneity of the older adult population. The manifestation and intensity of age-related changes can vary widely among individuals. The discrepancies between the pilots and the main study underscore the importance of adapting the study design with real-time feedback and participant needs, ensuring a more comfortable and meaningful experience for older participants.

\subsection{Data Collection}
\label{sec:study-measures}
We collected participants' responses to the pre-study questionnaire, in which we asked for information about their age, gender, education level, primary occupation, dominant hand, vision deficits, vision corrections, and color vision deficiency. We also collected information about their level of familiarity with visualization, familiarity with \baricon, \donuticon, and \radialicon chart, years of experience with visualization, and their self-reported tech-savviness (see supplementary material for complete questionnaire).

For each trial within a staircase, we collected the following data: exposure time in milliseconds, chart type and data size of the stimulus, and the correctness of the participant's answer to the comparison task. We also asked participants to ``think aloud,'' recording audio of and transcribing their thoughts and reactions during the study. Between sets of three staircases (i.\,e., three data sizes for a chart), participants were asked what strategy (if any) they used to perform the comparison task.

After completing the final staircase, each participant answered a post-study questionnaire. Participants were prompted to rank the three types of charts on a scale from 1 (best) to 3 (worst) based on their subjective preferences and confidence in accurately interpreting the data. We also inquired about their smartwatch ownership. For smartwatch owners, we asked about the watch's make, what they use it for, how often they monitor their data, and any visualizations they encounter when using it. We further explored the data types they actively monitor, additional data they would like to track, and their preference for visualizations over numerical displays. For those without a smartwatch, we gauged their interest in owning one and inquired about any barriers preventing them. Finally, all participants were shown a list of commonly tracked smartwatch data types identified by Islam et al.~\cite{islam2020visualizing} and asked to indicate which they would be interested in monitoring (see supplementary material for complete questionnaire).
\section{Data Analysis and Results}
This section is organized into five subsections:  Time Thresholds (\autoref{sec:results:thresholds}), Accuracy (\autoref{sec:results:accuracy}), Strategies (\autoref{sec:results:strategies}), Preference and Confidence Ratings (\autoref{sec:results:post-study}), and Smartwatch Ownership and Data of Interest (\autoref{sec:results:sw-own-data}).

\subsection{Time Thresholds}\label{sec:results:thresholds}
To gain a comprehensive understanding of the relationship between chronological age and time thresholds, we divided our analysis into three parts: (1) a between-groups comparison of means between younger adults (original study~\cite{blascheck2018glanceable}) and older adults (current study), (2) a between-groups comparison of means between young-old (age 65-74) and old-old (age $\geq 75$) adults from within our older adults group, and (3) the overall performance of older adults.

As per the original study, we created 95\% confidence intervals (CIs) of participant thresholds using BCa bootstrapping~\cite{efron1987better} per condition, using Bonferroni correction when making multiple comparisons (i.\,e., \autoref{fig:condition_comparisons}). All the information regarding the \textit{younger adults'} performance and preference we used for analysis in this section was initially collected and reported by Blascheck et al.~\cite{blascheck2018glanceable}.

To compare independent groups, younger vs. older adults as well as young-old vs. old-old, we assessed the strength of their time threshold differences for a specific condition by examining the overlap of their confidence intervals (CIs). In line with Blascheck et al.~\cite{blascheck2018glanceable}, we quantified this overlap with a metric we refer to as \textit{interval overlap percentage} (IOP), calculating and interpreting IOP following the guidelines provided by Cumming~\cite{cumming2014new} as well as Besançon and Dragicevic~\cite{besancon2017difference}; further details on this calculation can be found in the supplementary materials.

\subsubsection{Older Adults Combined}
\begin{figure*}
    \centering
    \begin{subfigure}{\textwidth}
        \centering
        \includegraphics[width=\linewidth]{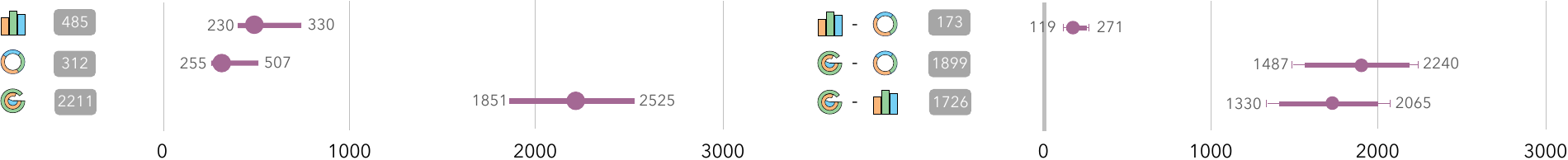}
        \caption{}
        \Description{Set of visualizations of confidence intervals for mean threshold times across all data sizes and older adult participants; tabular version provided in supplemental materials. Confidence intervals for each of the three chart types are on the left, and pairwise differences between all three chart types are on the right (with added thinner bars representing Bonferonni correction). The X-axis ranges from 0 to 3000~ms for both left and right, with chart confidence intervals stacked vertically. On the left, the mean for Radial charts is noticeably higher than the other two chart types, while Bar and Donut charts are close to each other. Also, the confidence interval width for Radial is larger than the other two chart types. On the right, the pairwise difference between Bar and Donut is much smaller, with the other two pairwise differences close to each other. Bar and Donut's interval width is much smaller than the other differences’ intervals.}
        \label{fig:conditions_breakdwon}
    \end{subfigure}
    
    \vspace{0.5em} 
    
    \begin{subfigure}{\textwidth}
        \centering
        \includegraphics[width=\linewidth]{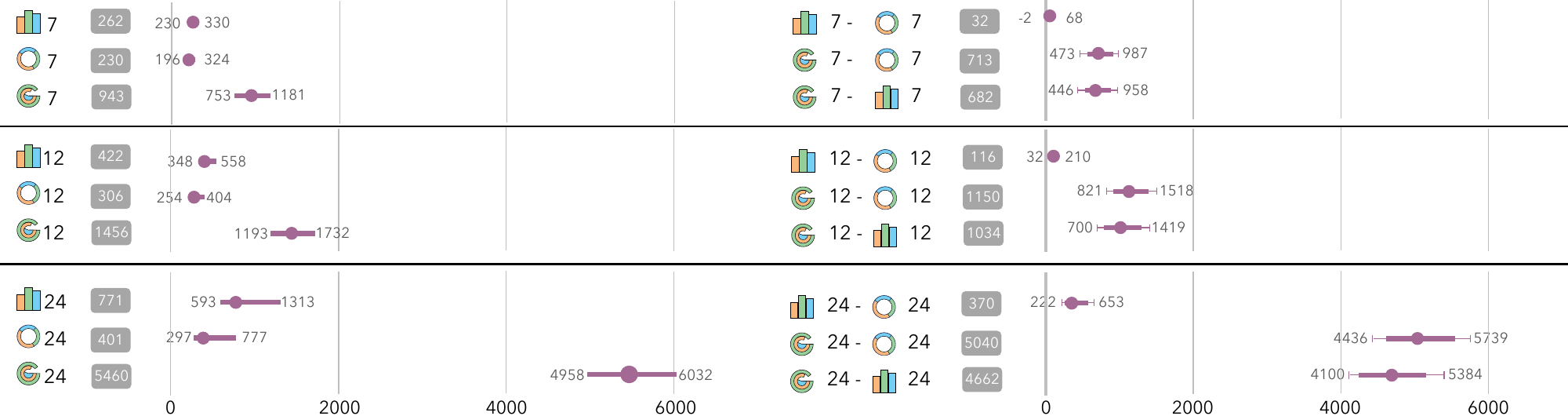}
        \caption{} 
        \Description{Set of confidence interval visualizations for mean threshold times for each of the data sizes across all older adult participants; tabular version provided in supplemental materials. Confidence intervals for each of the three chart types are on the left, and pairwise differences between all three chart types are on the right (with added thinner bars representing Bonferonni correction). The X-axis ranges from 0 to 6000~ms for both left and right, with chart confidence intervals stacked vertically. On the left, mean thresholds for the Radial condition are the highest for each data size, with the largest distance between chart types occurring for 24 data points. The confidence interval width for the Radial conditions is larger than the other two for all three data sizes. On the right, the pairwise difference between Radial and Donut is the lowest, with the largest distance between pairwise differences occurring for 24 data points. The confidence interval width Radial and Donut is much smaller than the other differences’ intervals for all three data sizes.}
        \label{fig:condition_comparisons}
    \end{subfigure}
    \vspace{-2em}
    \caption{(a-left) Average (\meanicon) time thresholds of older adults' (age $\geq 65$) performance per visualization type. (a-right) The pair-wise difference between the three visualization types. (b-left) Breakdown of older adults' time thresholds by visualization type and data size; original CIs are thicker lines, and Bonferroni corrections are thinner. (b-right) Pair-wise performance comparisons, articulated by visualization type and data size.}
    \label{fig:older_adults_combined}
\end{figure*}

This section delves into an in-depth analysis of the performance of older adults. In \autoref{fig:conditions_breakdwon}, the average time thresholds per visualization type (left)  and their pair-wise comparisons (right) are depicted. Older adults exhibited the best overall performance using \donuticon, with \baricon trailing closely behind. However, the performance gap notably widens for the \radialicon. For pair-wise comparisons (\autoref{fig:conditions_breakdwon}, right), the difference in performance between \donuticon and \baricon is considerably smaller than the difference between \donuticon and \radialicon or \baricon and \radialicon. These findings align well with the observations from the original study by Blascheck et al.~\cite{blascheck2018glanceable}. Despite the noticeable variances in time thresholds, the overarching trend suggests that \donuticon is most apt for facilitating quick comparison tasks across both demographics. \autoref{fig:condition_comparisons} (left) shows the breakdown of the older adults' performance by the nine experimental conditions, and \autoref{fig:condition_comparisons} (right) shows their pair-wise comparisons. Consistent with our following results, performance declined as the visualization complexity grew in conjunction with data size, with performance most notably affected for \radialicon. We also evaluated the impact of demographic factors (education level, technological familiarity, overall visualization familiarity, individual visualization familiarity, and smartwatch ownership) on participants' performance, finding no consistent patterns of differences based on our study population; more details can be found in the supplementary materials.

\subsubsection{Younger vs. Older Adults}\label{sec:younger-vs-older}
\begin{figure*}
    \centering
    \includegraphics[width=\textwidth]{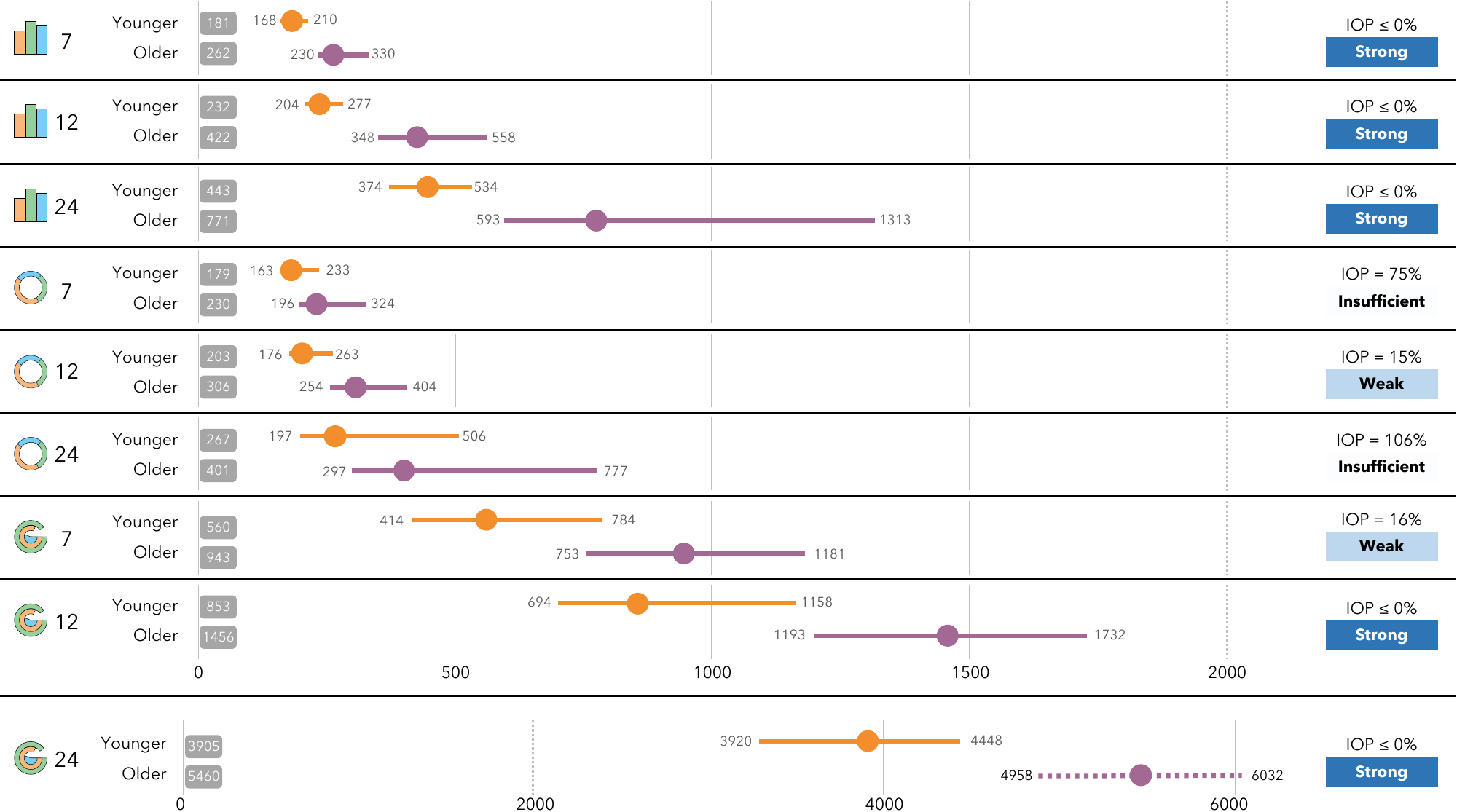}
    \caption{Mean time thresholds for younger~\yadultcbox (age $<65$) and older~\oadultcbox (age $\geq 65$) participants.~\meanicon shows the mean time thresholds. Note that, for \radialicon~24, only 17 of the older adults could complete the task. Additionally, \radialicon~24 ranges from $0$ to $6000$, whereas all other charts range from $0$ to $2000$.}
    \Description{Set of confidence interval visualizations for mean threshold times of each condition (chart and data size) for both younger adults (previous study) and older adults (current study); tabular version provided in supplemental materials. For all conditions except Radial 24, the X-axis ranges from 0 to 2000~ms. The older adults’ interval has a higher mean for each condition than younger adults. A thicker black line separates the Radial 24 condition since not all older adult participants completed that condition, and the X-axis ranges from 0 to 6000~ms.}
    \label{fig:older-younger-diffs}
\end{figure*}

\autoref{fig:older-younger-diffs} shows the average time threshold, CIs, and IOPs for younger and older adults, organized by the nine experimental conditions. At a high level, we observed that the average time thresholds for both younger and older adults increased in unison with data size (7 $\rightarrow$ 12 $\rightarrow$ 24) across all conditions. Younger adults seemed to perform consistently better than older adults across all conditions. We used the confidence interval overlap method to assess the magnitude of the observed differences between the two groups. The results showed \textbf{strong} evidence of differences for \baricon across all data sizes and \radialicon for 12 and 24 data points. We also found \textbf{weak} evidence of differences for \donuticon 12 and \radialicon 7. There was \textbf{insufficient} evidence of a difference for \donuticon 7 and \donuticon 24.

\begin{figure*}
    \centering
    \includegraphics[width=\textwidth]{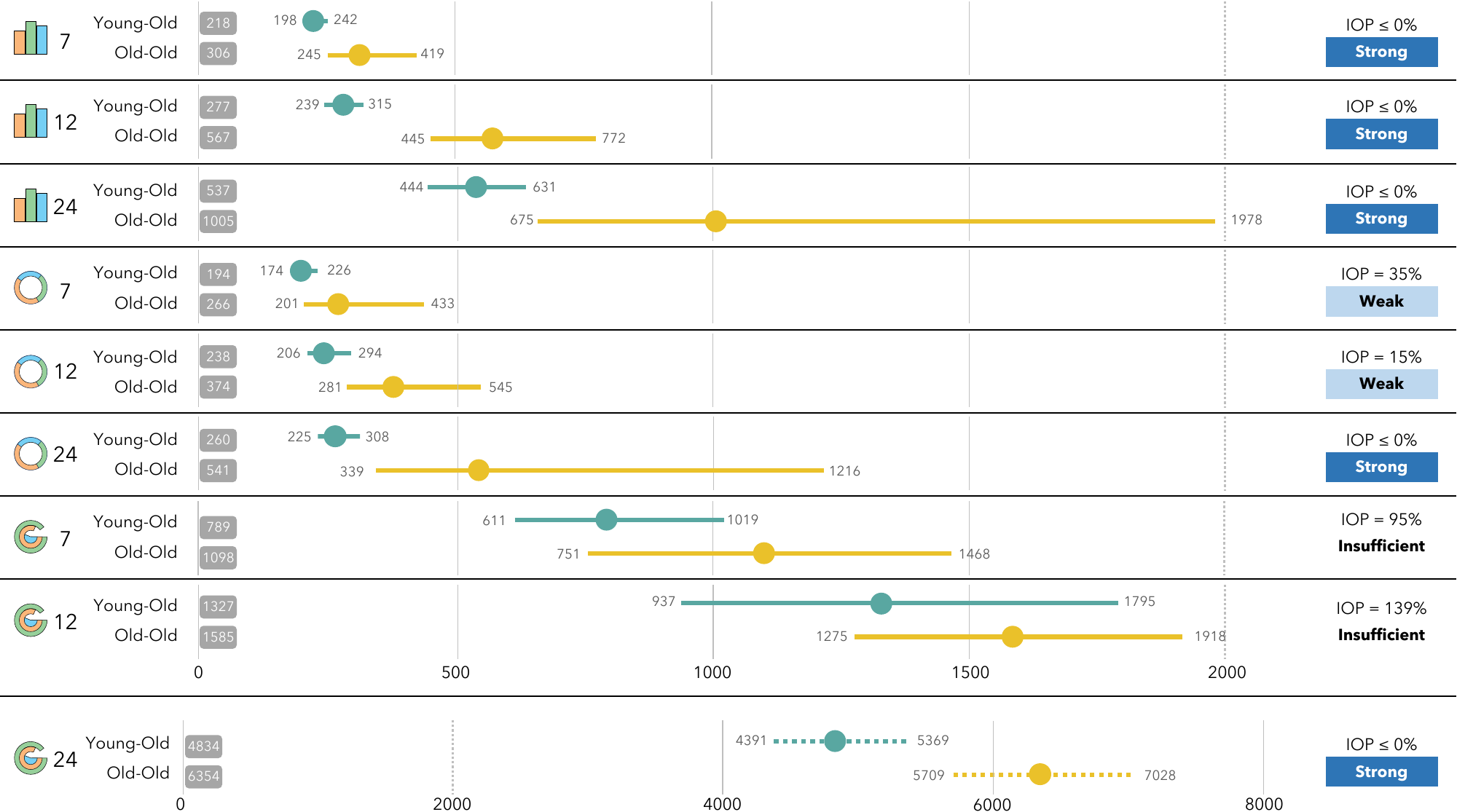}
    \caption{Mean time thresholds for young-old~\yoldcbox (age $65$-$74$) participants and old-old~\oldcbox (age $\geq 75$) participants.~\meanicon shows the mean time thresholds. Note that, for \radialicon~24, only 10 young-old and 7 old-old participants completed the task. Additionally, \radialicon~24 ranges from $0$ to $8000$, whereas all other charts range from $0$ to $2000$.}
    \Description{Set of confidence interval visualizations for mean threshold times of each condition (chart and data size) for both young-old (65 to 74) and old-old (75 plus) adults in the current study; tabular version provided in supplemental materials. For all conditions except Radial 24, the X-axis ranges from 0 to 2000~ms. For each condition, the old-old participants’ confidence interval has a higher mean than younger adults. A thicker black line separates the Radial 24 condition since not all older adult participants completed that condition, and the X-axis ranges from 0 to 8000~ms.}
    \label{fig:65-74-75-plus-diffs}
\end{figure*}

\subsubsection{Young-Old vs. Old-Old}\label{section:young-old-vs-old-old}
The comparative analysis between younger and older adults showed a variety of disparities in their performance levels. Subsequently, we narrowed our focus to older adults, aiming to discern whether the shifts in performance would amplify with advancing age. To this end, we compared the performance of the young-old (65-74) and old-old (age $\geq 75$) participants. \autoref{fig:65-74-75-plus-diffs} presents the average time thresholds, CIs, and IOPs for the two groups organized by the nine conditions. We found \textbf{strong} evidence of performance differences between the two groups for \baricon across all data sizes, \donuticon 24, and \radialicon 24. There was also \textbf{weak} evidence of differences in the \donuticon for 7 and 12 data points, and \textbf{insufficient} evidence was found for differences in the \radialicon for 7 and 12 data points. 

\begin{figure*}
    \centering
    \includegraphics[width=\textwidth]{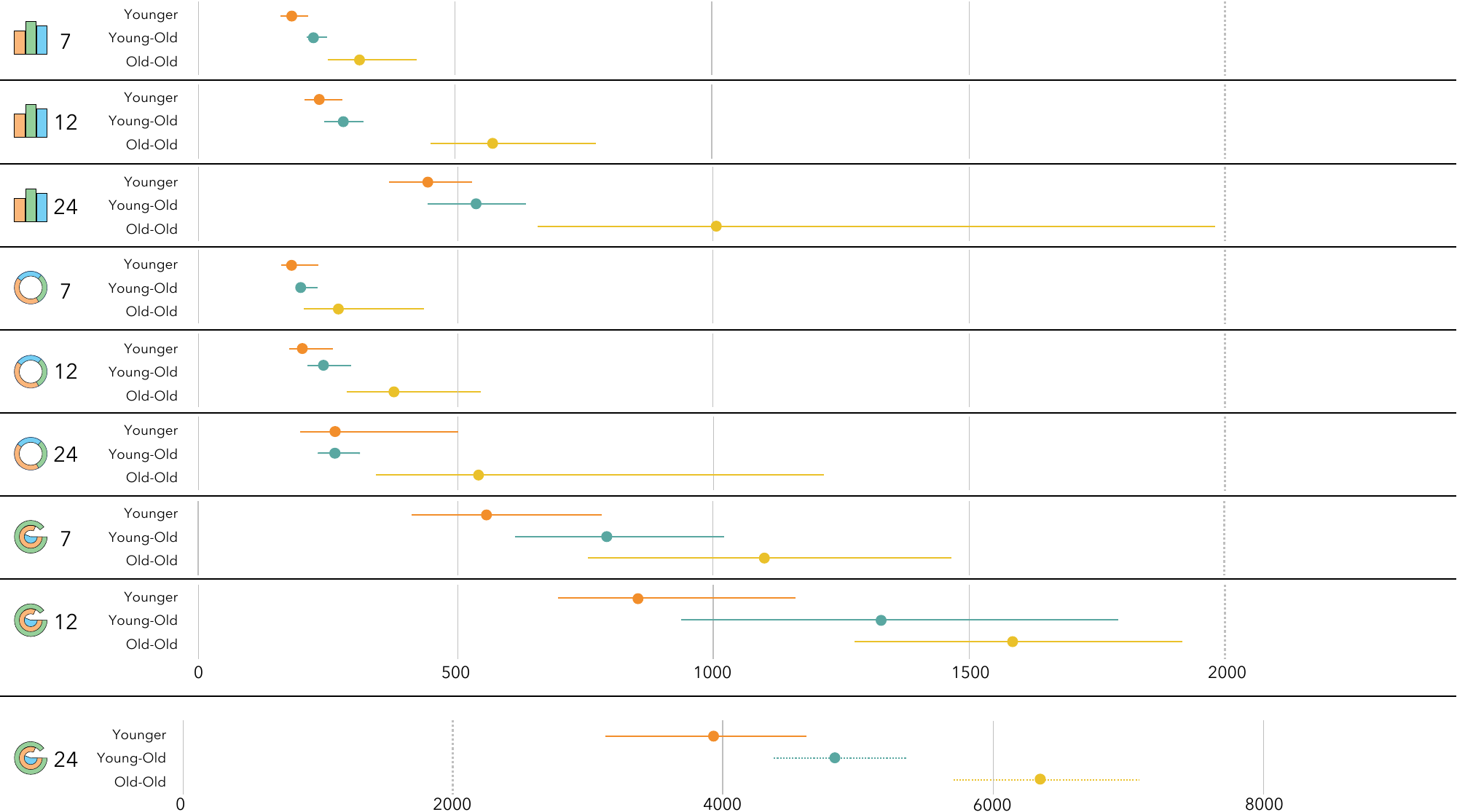}
    \caption{Mean time thresholds for younger~\yadultcbox (age $< 65$), young-old~\yoldcbox (age $65$-$74$), and old-old~\oldcbox (age $\geq 75$) adults. The discrepancies in younger and young-old are much less pronounced than the younger and the old-old. While age impacts time performance, the effects become more substantial as age advances.}\vspace{-1em}
    \Description{Set of confidence interval visualizations for mean threshold times of each condition (chart and data size), for younger adults from the previous study as well as young-old (65 to 74) and old-old (75 plus) adults in the current study; tabular version provided in supplemental materials. For all conditions except Radial 24, the X axis ranges from 0 to 2000~ms, while it ranges from 0 to 8000 for the Radial 24 condition. Overall, young adults and young-old adults are much closer to each other (similar means, lower values), while old-old adults have higher values. This pattern is less apparent for the three Radial conditions.}
    \label{fig:younger-65-74-older-75-plus-diffs}
\end{figure*}

These findings suggest a positive relationship between the decline in performance and advancing age. Hence, we next compared the performance of young-old and old-old participants against the younger adults, as illustrated in \autoref{fig:younger-65-74-older-75-plus-diffs}. While younger adults outperformed young-old adults, the performance discrepancies seemed relatively minor. In contrast, the performance gap between younger adults and the old-old (age $\geq 75$) was notably wider. We calculated and analyzed IOP among the three groups based on these observations. \autoref{tbl:yold-vs-oold-iop} shows the results of this analysis. We did not find strong evidence of a difference between the younger adults and the young-old. We only found \textbf{weak} evidence for \radialicon~24 and \textbf{insufficient} evidence for the eight remaining conditions. On the contrary, the evidence of differences between younger adults and the old-old was consistently strong apart from the differences for \radialicon~7 (weak) and \donuticon~24 (insufficient). The overall results of these analyses suggest that by and large, time thresholds appear to increase with age. However, the relationship between the two variables might be non-linear, and the rate of performance decline may accelerate with advancing age. 

\begin{table*}
    \centering
    \caption{The results of confidence interval overlap percentage (IOP) assessments between the younger adults (age $< 65)$ and the young-old (age 65-74) (left) as well as the younger adults and the old-old (age $\geq 75$) (right). The impact of aging on graphical perception appears to increase with the advancing of age.}\vspace{-1em}
    \label{tbl:yold-vs-oold-iop}
    \includegraphics[width=\textwidth]{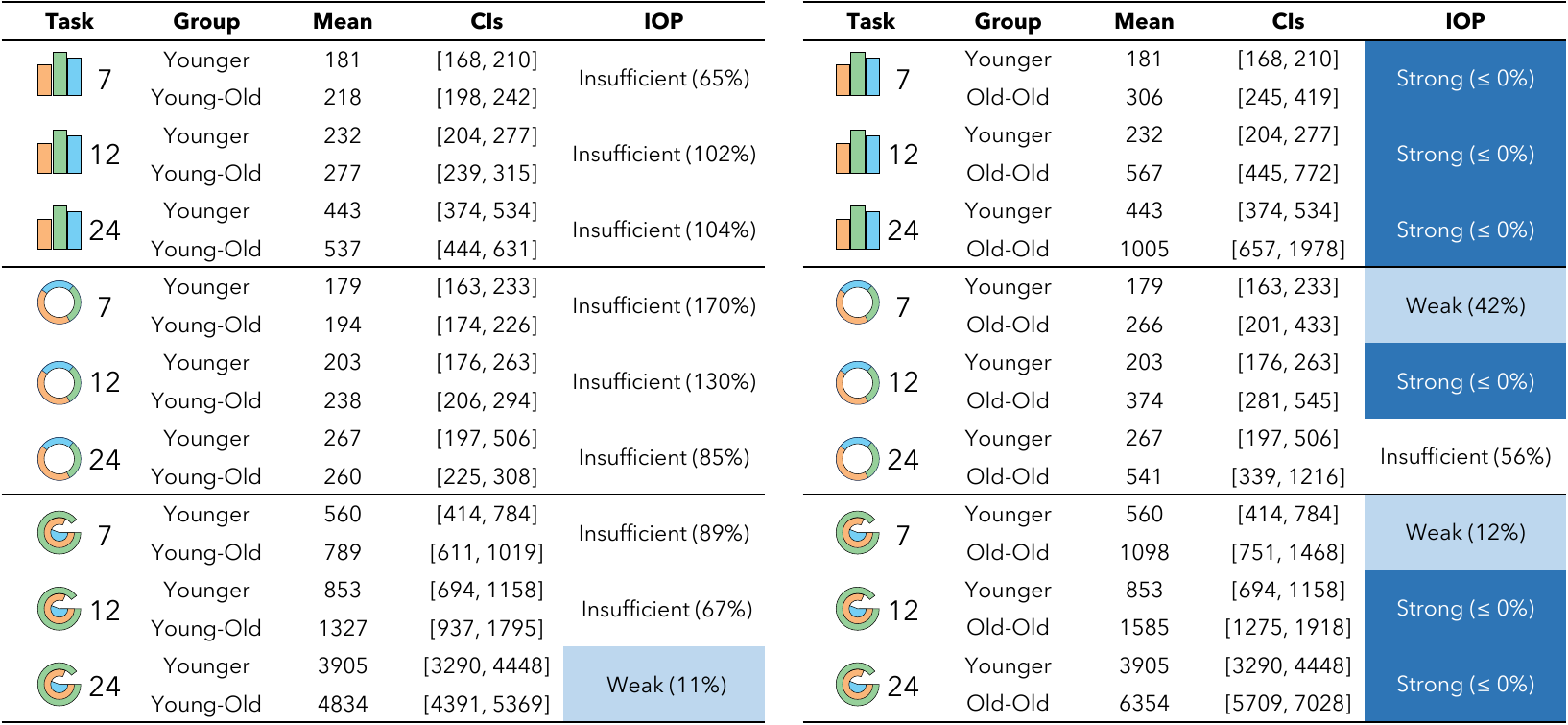}
    \label{tab:able_young-old_vs_old-old_IOPs}
\end{table*}

\begin{table}
\caption{Average accuracy for young-old (age 65-74) and old-old (age $\geq 75$) participants for each condition (chart type and data size).}\vspace{-1em}
\resizebox{\columnwidth}{!}{
\begin{tabular}{ccccccccc}
\hline
\textbf{} & \multicolumn{2}{c}{\baricon} &  & \multicolumn{2}{c}{\donuticon} &  & \multicolumn{2}{c}{\radialicon} \\
\textbf{Data Size} & \textbf{Young-Old} & \textbf{Old-Old} &  & \textbf{Young-Old} & \textbf{Old-Old} &  & \textbf{Young-Old} & \textbf{Old-Old} \\ \hline
7 & 77\% & 75\% &  & 79\% & 77\% &  & 68\% & 68\% \\
12 & 71\% & 72\% &  & 77\% & 74\% &  & 64\% & 63\% \\
24 & 69\% & 68\% &  & 74\% & 70\% &  & 63\% & 56\% \\ \hline
\end{tabular}%
}
\label{tab:accuracy}
\end{table}

\subsection{Accuracy}\label{sec:results:accuracy}

We assessed participants' performance by examining the accuracy of their responses, aiming to determine whether observed accuracy aligned closely with the anticipated accuracy of $\sim$63\%, as specified by the staircase design (refer to \autoref{sec:task-and-stimuli} for details). To achieve this, we calculated the average accuracy for each participant across each of the nine staircases they completed, dividing the number of correct answers by the total trials per staircase. \autoref{tab:accuracy} provides this analysis, segmented by the young-old and the old-old cohorts. Our study revealed that, except for old-old participants in the \radialicon 24 condition, both groups' accuracy matched or exceeded the expected value of $\sim$63\%.

\subsection{Strategies}\label{sec:results:strategies}
After finishing three consecutive staircases for the same chart type (7, 12, and 24 data points), we asked participants to describe the strategy they used for the task. This section describes all strategies per chart that more than one participant reported. It is important to note that some participants mentioned multiple strategies. 

\subsubsection{\baricon}
One strategy (8/24) revolved around identifying only one target bar and estimating based on that bar's size, for example, if the bar was short, then concluding that the other is likely larger. Another strategy (7/24) involved identifying both target bars and comparing their heights. Participants also mentioned a third strategy, in which they answered based on the overall shape of the data distribution (6/24) or the shapes of local distributions around the target bars (5/24). 

\subsubsection{\donuticon}
Many participants (13/24) described a strategy of taking in a single view of the entire chart or focusing on the center of the chart, either approximating the distribution or quickly looking peripherally for the larger patch of color with a dot in it; this was mentioned to work well except in cases where the target elements were similar in size or small. Identical to \baricon, some participants (8/24) would focus on only one element and would estimate based on its size. Other participants (6/24) would try to find both dots and compare the regions if given enough time, which was mentioned to be especially helpful for small and similarly-sized regions.

\subsubsection{\radialicon}
Participants most commonly (12/24) reported giving answers based on only focusing on the inner target bar and estimating its value compared to the other target. Some participants (4/24) compared clusters of bars that included the targets, while others (2/24) tried to check each dot and follow its arc.

\subsection{Preference and Confidence Ratings}
\label{sec:results:post-study}
For each data size, participants ranked the three chart types in terms of their subjective preferences and confidence in accurately interpreting the data. \autoref{tbl:pref-ranks} shows the responses of younger (prior study) and older adults normalized to percentages of their respective sample sizes. We also break down the older adults group into the young-old and old-old subsets, presenting their preference and confidence ratings in \autoref{tbl:pref-ranks-yooo}.

Across all data sizes, older adults had the greatest preference for and confidence with \donuticon, followed by \baricon and \radialicon. 
While the percentages of older adults who ranked \baricon and \donuticon the highest were similar for data size 7 (25~pp difference for preference, 16~pp for confidence), the differences increased with the data size, resulting in a more than 50~pp discrepancy for 12 and 24.

For preference, older adults ranked \donuticon the highest across all data sizes, while younger adults ranked \baricon the highest for data size 7 and \donuticon for data sizes 12 and 24. For \baricon 12, 66~pp more of older adults rated \donuticon as the highest compared to \baricon with a difference of only 12~pp for younger adults. The Rank 1 differences for data size 24 were slightly closer, with a 66~pp difference between \donuticon and \baricon for older adults and 34~pp for younger adults. For confidence, older and younger adults had similar results (\donuticon $>$ \baricon $>$ \radialicon) for all data sizes.

The results for the young-old and old-old are similar for preference and confidence. The largest difference for preference is for Rank 1 of \baricon with 12 data points (25~pp) and 24 data points (25~pp). In contrast, the greatest difference in confidence is 25~pp for both Rank 1 and Rank 2 of \baricon 12 and \donuticon 12.

\newcommand{\pctwid}{0.65cm}
\begin{table}[t]
\centering
\footnotesize
    \caption{Preference and confidence ratings for younger (age $< 65$) (Y) and older (age $\geq 65$) (O) adults, per chart and data size (DS). Because each sample had a different size ($n_Y=18,n_O=24$), percentages of each sample are instead presented.}\vspace{-1em}
    \label{tbl:pref-ranks}
\resizebox{0.49\textwidth}{!}{%
\begin{tabular}{ccp{\pctwid}p{\pctwid}p{\pctwid}p{\pctwid}p{\pctwid}p{\pctwid}}
\hline
\multicolumn{8}{c}{\textbf{Preference}} \\ \hline
& & \multicolumn{2}{c}{\baricon}& \multicolumn{2}{c}{\donuticon}& \multicolumn{2}{c}{\radialicon} \\
\textbf{DS}&\textbf{Rank}&\textbf{\% Y} &\textbf{\% O}&\textbf{\% Y} &\textbf{\% O}&\textbf{\% Y} &\textbf{\% O} \\ \hline
\multirow{3}{*}{7} & 1 & \databar[donutyoung]{56} & \databar[donutyoung]{38}& \databar[donutyoung]{44} & \databar[donutyoung]{63}& \databar[donutyoung]{0} & \databar[donutyoung]{0} \\
 & 2 & \databar[donutyoung]{33} & \databar[donutyoung]{58}& \databar[donutyoung]{56} & \databar[donutyoung]{38}& \databar[donutyoung]{11} & \databar[donutyoung]{4} \\
 & 3 & \databar[donutyoung]{11} & \databar[donutyoung]{4}& \databar[donutyoung]{0} & \databar[donutyoung]{0}& \databar[donutyoung]{89} & \databar[donutyoung]{96} \\ \hline
\multirow{3}{*}{12} & 1 & \databar[donutyoung]{44} & \databar[donutyoung]{13}& \databar[donutyoung]{56} & \databar[donutyoung]{79}& \databar[donutyoung]{0} & \databar[donutyoung]{8} \\
 & 2 & \databar[donutyoung]{44} & \databar[donutyoung]{79}& \databar[donutyoung]{44} & \databar[donutyoung]{21}& \databar[donutyoung]{11} & \databar[donutyoung]{0} \\
 & 3 & \databar[donutyoung]{11} & \databar[donutyoung]{8}& \databar[donutyoung]{0} & \databar[donutyoung]{0}& \databar[donutyoung]{89} & \databar[donutyoung]{92} \\ \hline
\multirow{3}{*}{24} & 1 & \databar[donutyoung]{33} & \databar[donutyoung]{13}& \databar[donutyoung]{67} & \databar[donutyoung]{79}& \databar[donutyoung]{0} & \databar[donutyoung]{8} \\
 & 2 & \databar[donutyoung]{61} & \databar[donutyoung]{79}& \databar[donutyoung]{33} & \databar[donutyoung]{21}& \databar[donutyoung]{6} & \databar[donutyoung]{0} \\
 & 3 & \databar[donutyoung]{6} & \databar[donutyoung]{8}& \databar[donutyoung]{0} & \databar[donutyoung]{0}& \databar[donutyoung]{94} & \databar[donutyoung]{92} \\ \hline
\end{tabular}%
}
\hfill
\resizebox{0.49\textwidth}{!}{%
\begin{tabular}{ccp{\pctwid}p{\pctwid}p{\pctwid}p{\pctwid}p{\pctwid}p{\pctwid}}
\hline
\multicolumn{8}{c}{\textbf{Confidence}} \\ \hline
&& \multicolumn{2}{c}{\baricon}& \multicolumn{2}{c}{\donuticon}& \multicolumn{2}{c}{\radialicon} \\
\textbf{DS}&\textbf{Rank}&\textbf{\% Y} &\textbf{\% O}&\textbf{\% Y} &\textbf{\% O}&\textbf{\% Y} &\textbf{\% O} \\ \hline
\multirow{3}{*}{7} & 1 & \databar[donutyoung]{39} & \databar[donutyoung]{42}& \databar[donutyoung]{61} & \databar[donutyoung]{58}& \databar[donutyoung]{0} & \databar[donutyoung]{0} \\
 & 2 & \databar[donutyoung]{56} & \databar[donutyoung]{58}& \databar[donutyoung]{39} & \databar[donutyoung]{42}& \databar[donutyoung]{6} & \databar[donutyoung]{0} \\
 & 3 & \databar[donutyoung]{6} & \databar[donutyoung]{0}& \databar[donutyoung]{0} & \databar[donutyoung]{0}& \databar[donutyoung]{94} & \databar[donutyoung]{100} \\ \hline
\multirow{3}{*}{12} & 1 & \databar[donutyoung]{22} & \databar[donutyoung]{13}& \databar[donutyoung]{78} & \databar[donutyoung]{88}& \databar[donutyoung]{0} & \databar[donutyoung]{0} \\
 & 2 & \databar[donutyoung]{72} & \databar[donutyoung]{88}& \databar[donutyoung]{22} & \databar[donutyoung]{13}& \databar[donutyoung]{11} & \databar[donutyoung]{0} \\
 & 3 & \databar[donutyoung]{6} & \databar[donutyoung]{0}& \databar[donutyoung]{0} & \databar[donutyoung]{0}& \databar[donutyoung]{89} & \databar[donutyoung]{100} \\ \hline
\multirow{3}{*}{24} & 1 & \databar[donutyoung]{11} & \databar[donutyoung]{17}& \databar[donutyoung]{89} & \databar[donutyoung]{83}& \databar[donutyoung]{0} & \databar[donutyoung]{0} \\
 & 2 & \databar[donutyoung]{78} & \databar[donutyoung]{79}& \databar[donutyoung]{11} & \databar[donutyoung]{17}& \databar[donutyoung]{11} & \databar[donutyoung]{4} \\
 & 3 & \databar[donutyoung]{6} & \databar[donutyoung]{4}& \databar[donutyoung]{0} & \databar[donutyoung]{0}& \databar[donutyoung]{89} & \databar[donutyoung]{96} \\ \hline
\end{tabular}%
}
\end{table}

\begin{table}[t]
\centering
\footnotesize
    \caption{Preference and confidence ratings for young-old (age 65-74) (YO) and old-old (age $\geq 75$) (OO) adults, per chart and data size (DS).}\vspace{-1em}
    \label{tbl:pref-ranks-yooo}
\resizebox{0.49\textwidth}{!}{%
\begin{tabular}{ccp{\pctwid}p{\pctwid}p{\pctwid}p{\pctwid}p{\pctwid}p{\pctwid}}
\hline
\multicolumn{8}{c}{\textbf{Preference}} \\ \hline
& & \multicolumn{2}{c}{\baricon} & \multicolumn{2}{c}{\donuticon} & \multicolumn{2}{c}{\radialicon} \\
\textbf{DS}&\textbf{Rank} &\textbf{\% YO} &\textbf{\% OO} &\textbf{\% YO} &\textbf{\% OO} &\textbf{\% YO} &\textbf{\% OO} \\ \hline
\multirow{3}{*}{7} & 1 & \databar[donutyoung]{33} & \databar[donutyoung]{42} & \databar[donutyoung]{67} & \databar[donutyoung]{58} & \databar[donutyoung]{0} & \databar[donutyoung]{0} \\
 & 2 & \databar[donutyoung]{58} & \databar[donutyoung]{58} & \databar[donutyoung]{33} & \databar[donutyoung]{42} & \databar[donutyoung]{8} & \databar[donutyoung]{0} \\
 & 3 & \databar[donutyoung]{8} & \databar[donutyoung]{0} & \databar[donutyoung]{0} & \databar[donutyoung]{0} & \databar[donutyoung]{92} & \databar[donutyoung]{100} \\ \hline
\multirow{3}{*}{12} & 1 & \databar[donutyoung]{0} & \databar[donutyoung]{25} & \databar[donutyoung]{83} & \databar[donutyoung]{75} & \databar[donutyoung]{17} & \databar[donutyoung]{0} \\
 & 2 & \databar[donutyoung]{83} & \databar[donutyoung]{75} & \databar[donutyoung]{17} & \databar[donutyoung]{25} & \databar[donutyoung]{0} & \databar[donutyoung]{0} \\
 & 3 & \databar[donutyoung]{17} & \databar[donutyoung]{0} & \databar[donutyoung]{0} & \databar[donutyoung]{0} & \databar[donutyoung]{83} & \databar[donutyoung]{100} \\ \hline
\multirow{3}{*}{24} & 1 & \databar[donutyoung]{0} & \databar[donutyoung]{25} & \databar[donutyoung]{83} & \databar[donutyoung]{75} & \databar[donutyoung]{17} & \databar[donutyoung]{0} \\
 & 2 & \databar[donutyoung]{83} & \databar[donutyoung]{75} & \databar[donutyoung]{17} & \databar[donutyoung]{25} & \databar[donutyoung]{0} & \databar[donutyoung]{0} \\
 & 3 & \databar[donutyoung]{17} & \databar[donutyoung]{0} & \databar[donutyoung]{0} & \databar[donutyoung]{0} & \databar[donutyoung]{83} & \databar[donutyoung]{100} \\ \hline
\end{tabular}%
}
\hfill
\resizebox{0.49\textwidth}{!}{%
\begin{tabular}{ccp{\pctwid}p{\pctwid}p{\pctwid}p{\pctwid}p{\pctwid}p{\pctwid}}
\hline
\multicolumn{8}{c}{\textbf{Confidence}} \\ \hline
& & \multicolumn{2}{c}{\baricon} & \multicolumn{2}{c}{\donuticon} & \multicolumn{2}{c}{\radialicon} \\
\textbf{DS}&\textbf{Rank} &\textbf{\% YO} &\textbf{\% OO} &\textbf{\% YO} &\textbf{\% OO} &\textbf{\% YO} &\textbf{\% OO} \\ \hline
\multirow{3}{*}{7} & 1 & \databar[donutyoung]{33} & \databar[donutyoung]{50} & \databar[donutyoung]{67} & \databar[donutyoung]{50} & \databar[donutyoung]{0} & \databar[donutyoung]{0} \\
 & 2 & \databar[donutyoung]{67} & \databar[donutyoung]{50} & \databar[donutyoung]{33} & \databar[donutyoung]{50} & \databar[donutyoung]{0} & \databar[donutyoung]{0} \\
 & 3 & \databar[donutyoung]{0} & \databar[donutyoung]{0} & \databar[donutyoung]{0} & \databar[donutyoung]{0} & \databar[donutyoung]{100} & \databar[donutyoung]{100} \\ \hline
\multirow{3}{*}{12} & 1 & \databar[donutyoung]{0} & \databar[donutyoung]{25} & \databar[donutyoung]{100} & \databar[donutyoung]{75} & \databar[donutyoung]{0} & \databar[donutyoung]{0} \\
 & 2 & \databar[donutyoung]{100} & \databar[donutyoung]{75} & \databar[donutyoung]{0} & \databar[donutyoung]{25} & \databar[donutyoung]{0} & \databar[donutyoung]{0} \\
 & 3 & \databar[donutyoung]{0} & \databar[donutyoung]{0} & \databar[donutyoung]{0} & \databar[donutyoung]{0} & \databar[donutyoung]{100} & \databar[donutyoung]{100} \\ \hline
\multirow{3}{*}{24} & 1 & \databar[donutyoung]{8} & \databar[donutyoung]{25} & \databar[donutyoung]{92} & \databar[donutyoung]{75} & \databar[donutyoung]{0} & \databar[donutyoung]{0} \\
 & 2 & \databar[donutyoung]{83} & \databar[donutyoung]{75} & \databar[donutyoung]{8} & \databar[donutyoung]{25} & \databar[donutyoung]{8} & \databar[donutyoung]{0} \\
 & 3 & \databar[donutyoung]{8} & \databar[donutyoung]{0} & \databar[donutyoung]{0} & \databar[donutyoung]{0} & \databar[donutyoung]{92} & \databar[donutyoung]{100} \\ \hline
\end{tabular}
}
\end{table}

\subsection{Smartwatch Ownership and Data of Interest}
\label{sec:results:sw-own-data}
Six of the twenty-four participants reported owning a smartwatch. They primarily used their devices for health (heart rate, oxygen, fall detection), activity (step count) monitoring, weather, and voice memos. The frequency of use varied from several times a day~(4/6) to very seldom~(1/6), with one stating that their frequency varied. Five of the six participants preferred viewing their data as numbers, with one preferring to see it presented as a visualization. The remaining eighteen participants did not own a smartwatch. Stated reasons for not owning a device included not needing it~(11/18), not wanting to track personal data~(7/18), assumptions of high cost~(6/18), a lack of interest in technology~(5/18), a lack of interest in wrist accessories~(5/18), not knowing what smartwatches do~(4/18), uncertainty with learning how to use it~(3/18), avoiding reliance on technology~(2/18), and preference for larger screens~(2/18). 

\begin{figure*}
    \centering
    \includegraphics[width=0.85\textwidth]{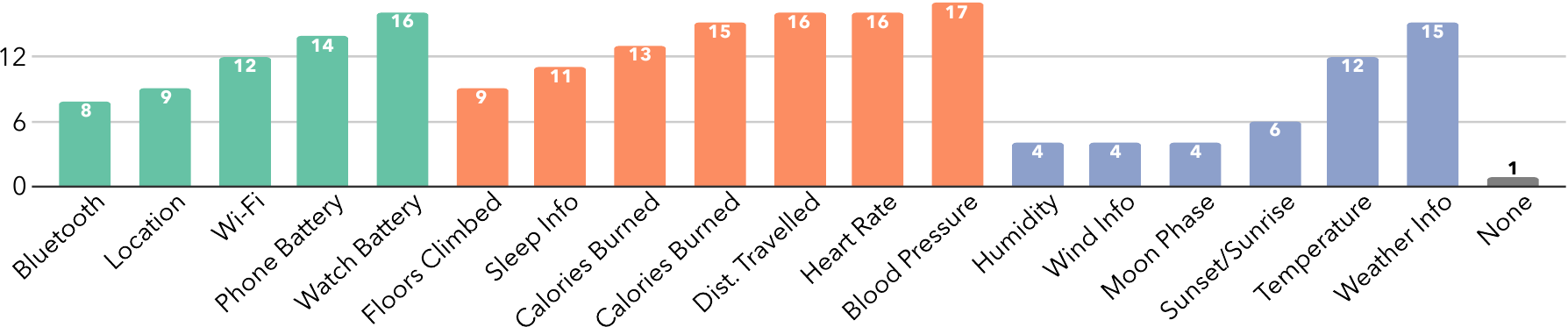}
    \vspace{-1em}
    \caption{Smartwatch data of interest for older adults (age $\geq 65$) in our study. Each bar represents the number of participants interested in that data type. Categories are \texttt{Device-Location}~\devicecbox, \texttt{Health-Fitness}~\healthcbox, and \texttt{Weather \& Planetary}~\weathercbox as per a previous study by Islam et al.~\cite{islam2020visualizing}. One participant in our study chose \textit{None}~\nonecbox as an answer.}
    \Description{Vertical bar chart depicting the number of older adults (current study participants) who said they would be interested in viewing various data types on a smartwatch; tabular version provided in supplemental materials. Data types are put into three possible categories: Device, Health, and Weather. In the Weather category, four data types were selected by 4 or fewer participants (humidity, wind info, moon phase, and sunset/sunrise). In comparison, the other two Weather data types were chosen by 12 or more people (temperature and weather info). A fourth category of None is shown, with 1 participant selecting that.}
    \label{fig:sw-data-interest}
\end{figure*}

In our study, participants were asked to select data types (as per Islam et al.~\cite{islam2020visualizing}) that they would be interested in tracking on a smartwatch. Of the three major categories, \texttt{Health-Fitness} had the highest overall interest, followed by \texttt{Device-Location} and \texttt{Weather \& Planetary} data. \textit{Blood Pressure} was the most favored type of data (17/24), while \textit{Humidity} (4/24), \textit{Wind Info} (4/24), \textit{Moon Phase} (4/24), and \textit{None of the Above} (1/24) were the least-selected (see \autoref{fig:sw-data-interest}). Compared with Islam et al.~\cite{islam2020visualizing} (in which participants indicated data they \textit{already see} on their smartwatch), older adults in our study showed a greater interest in \textit{Blood Pressure}, \textit{Distance Travelled}, and \textit{Phone Battery}. 

\section{Discussion}
\label{sec:discussion}
In this section, we discuss the findings of our work, highlighting differences and similarities between younger and older adults' performance, preferences, and confidence using glanceable visualizations. We explore the implications for visualization design and use in the real world. Next, we share our experiences and lessons learned from conducting a study with older adults, offering insights for researchers interested in working with similar populations. Lastly, we offer preliminary guidance for visualization designers working on glanceable visualizations for older adults.

\subsection{Aging Agility: Similar Trends With Slower Times}
\label{sec:aging-agility}

Across all experimental conditions, older adults' performance consistently lagged behind that of the younger participants from the original study~\cite{blascheck2018glanceable}. The differences in average time thresholds ranged from 51~ms (\donuticon~7) to 1555~ms (\radialicon~24), and except \donuticon~24, the performance gap invariably widened in tandem with increasing data size. The confidence interval overlap analysis showed strong (5/9) or weak (2/9) evidence of differences in seven out of nine conditions, as illustrated in \autoref{fig:older-younger-diffs}. Despite these disparities, the overarching trends related to visualization type, data size, and time thresholds were comparable between younger and older adults. Both groups achieved optimal performance with \donuticon, followed by \baricon and \radialicon. Also, in almost all cases, both studies' participants subjectively ranked \donuticon higher than \baricon and \radialicon. 

The findings of our study suggest that some aspects of graphical perception remain unchanged with age (e.\,g., preferences) while others (e.\,g., speed) may decline. The similarities mentioned above between younger and older adults might indicate that basic pattern recognition, which includes identifying basic geometric shapes, lines, orientations, and spatial relationships among visual elements, remains relatively stable with age~\cite{murman2015impact}. Older adults were, however, invariably slower than their younger counterparts. This may partially stem from the compounded effects of perceptual and cognitive changes that people may experience with age. For instance, each comparison task required older adults to search for small black dots that marked the target elements. Visual search, which is the process of actively scanning the visual scene to identify a specific target among distractors, becomes slower and more error-prone with age, especially as the complexity of the visual environment increases~\cite{becic2007age}. Comparing visual stimuli also relies on several cognitive functions, including working memory (WM), the mechanism responsible for the short-term simultaneous holding and manipulating information in our awareness. WM, however, declines with age, and on average, the WM of older adults is lower than that of their younger counterparts~\cite{salthouse1991decomposing}. Hence, we speculate that declining WM could decrease the speed and accuracy of performing comparisons; however, it is essential to note that our study neither controlled for nor measured the impact of WM; further work is needed to understand the effects of WM among different age groups in this context. Still, the considerable differences between the performance of the young-old and the old-old in our study (\autoref{fig:younger-65-74-older-75-plus-diffs}) suggest that the advancement of aging can exacerbate the performance disparities between older and younger adults. Due to our study design, the time taken to input the answer did not play a role in our measurements and can, therefore, be excluded as a possible source of differences. 

We found that older adults used various strategies to perform the comparison task. We speculate that one factor that impacted participants' strategies was the stimuli exposure time, based on remarks some participants made. For instance, P9 mentioned that \textit{``I tried to look at one [target bar]...and then the second dot [target bar], and see how tall it was. And that worked really well when things were slow. And then, when things were going faster...I tried to just kind of scan the bottom for the 2 dots and quickly look up and not focus on one versus the other, but try to see the proportion [distribution shape]''} and P18 said  \textit{``First I tried to see the whole thing. And when it got faster, and I couldn't do that because it was too fast to look, and I would miss it, so I would just look at one side and just guess to see if the other one was lower or higher than that one.''} Other factors, such as chart and data size, could have also influenced the strategies used to perform the task. However, further research is required to investigate these factors' combined and isolated effects on strategy.

An important question that emerges from the findings of our study is: \textit{``Does the slower performance of older adults in interacting with glanceable visualizations on smartwatches bear practical significance in real-world situations?''} Answering this question definitively necessitates further empirical examination. However, we posit that the relevance of slower performance may depend on a multifaceted interplay of factors, including the context, in which glancing occurs. Glimpsing at a smartwatch for quick insights often co-occurs with other activities, such as walking, biking, or daily chores~\cite{pizza2016smartwatch}. Older adults in the study reported a much greater interest in health data than younger adults in the survey by Islam et al.~\cite{islam2020visualizing}, and health markers are often tracked on a watch during exercise. In these dynamic contexts, the extended glance duration observed in older adults might be consequential, particularly in scenarios that demand continuous attention and precise coordination. For instance, when navigating complex terrain, the interplay of visual attention and exact timing is critical; even a brief distraction could enhance the risk of falls, especially among older adults who are walking~\cite{marigold2007gaze, yogev2008role}. The even slower performance of the old-old, particularly when compounded by more pronounced age-related physiological decline, might intensify such risks. Conversely, the additional time required to read a glanceable visualization might be negligible when an older adult is engaged in a less demanding context, such as sitting in a chair. Further studies must move beyond conjecture and deeply understand the factors influencing older adults' performance and experience with glanceable visualizations and their tangible real-world applications. We can extend the conversation from a specific empirical finding to a broader discourse on the relationship between visualization design and utility, aging, and the human experience.

\subsection{Conundrum of Chronology: The Challenge of Defining `Old' in Studies}
In line with the World Health Organization's definition of older adults, as well as prevalent research practices, we set the starting chronological age for recruiting participants in our study at 65. However, our breakdown analysis of the young-old (age 65-74) and old-old participants (age $\geq 75$) showed notable performance differences between the two groups (\autoref{fig:younger-65-74-older-75-plus-diffs}). This observation raises a critical question regarding visualization research involving older adults: \emph{``Who should we consider as older adults in the context of visualization research?''} The heterogeneity of the older adult population is a well-known phenomenon. While aging is a shared experience, the onset and intensity of age-related physiological changes differ among individuals~\cite{hofer2003evaluating,ylikoski1999heterogeneity}. Therefore, defining an age range that classifies older adults is challenging and debatable. We need mechanisms that enable us to enhance the internal and external validity of the experiments by recruiting samples that more accurately represent the older adult demographic. This might entail the creation of more systematic approaches to participant recruitment in visualization studies that assess core elements such as perception, cognition, and motor control. Pre-existing psychophysics exams like the mini-mental state test~\cite{schatz2011mini} could serve as blueprints for crafting analogous assessments in the field of visualization. In the absence of standardized protocols, and drawing from the findings of this study, we strongly recommend that researchers interested in studying older adults embrace a comprehensive age range in their recruitment, with particular emphasis on including an ample number of participants from the older and more advanced age categories. This is crucial to enhance the probability of observing and accurately measuring the impact of age-related changes in their studies. A failure to secure a representative population could increase the risk of \textit{type-I} and \textit{type-II} errors, leading to erroneous insights and conclusions. For instance, if we had solely recruited participants aged 60-74, we may have falsely inferred that no noteworthy differences exist between older and younger adults---an assertion contradicted by our age $\geq 75$ results.

\subsection{Methodological Flexibility: Dynamically Adapting to the Needs of Older Participants}
In the course of running this study, we made two adjustments to the study design and procedure. First, we allowed participants to exit the \radialicon~24 condition if performing trials became too cumbersome for them, and second, we replaced the keyboard with two separate keys (see \autoref{sec:study-design-exec} for details). Typically, changing the study protocol is not advised to preserve the integrity of the experimental conditions across participants. However, our experience in this study, and also evidence from prior work~\cite{moore1999initiating,roller2015applied}, suggest that methodological flexibility might be required when working with vulnerable populations. Participation in studies can cause anxiety in older adults for many reasons, such as misunderstanding the goals of the researcher, fatigue, and cognitive difficulty, which can lead to increased difficulty in finding participants~\cite{mchenry2015recruitment}. This recognition brings to light a critical consideration for researchers: the ethical imperative of balancing scientific rigor with human compassion and understanding. Hodge et al.~\cite{hodge2020relational} touch on a related topic, spotlighting the ethical issues that can arise with people while working in dementia research. It reminds us that study designs must not only be methodologically sound but also adaptable to the unique needs and challenges different populations face. It also opens up further research questions concerning the recruitment and retention of older adults. How can researchers establish trust and alleviate potential anxieties? What additional support mechanisms might be necessary to ensure the comfort and understanding of participants from varying age groups? These considerations emphasize the interconnected nature of research design, ethics, and the potential to contribute meaningfully to the lives of the individuals involved.

\subsection{Design Implications}
The primary objective of our study was to investigate and assess the graphical perception of older adults in the context of glanceable visualization. Drawing on our findings, we also offer preliminary suggestions to assist visualization designers working on glanceable visualizations for older adults.\\\vspace{-1em}

\noindent\textbf{\donuticon and \baricon are Preferable for Quick Value Comparisons.} The \donuticon and \baricon are comparably effective for supporting the comparison task for small data sets. In our study, older adults exhibited similar efficiency with both graph types for 7 data points, averaging a threshold difference of 32~ms. However, for larger data sets, older adults' performance with \donuticon notably outmatched their performance with \baricon, with average thresholds of 401~ms versus 771~ms for 24 data points. Blascheck et al. also found performance degradation in younger adults with larger datasets.
However, the average threshold difference between \donuticon and \baricon for older adults was approximately twice that of younger adults for each data size. While younger adults' average thresholds were less than 500~ms for both \baricon and \donuticon with 24 data points, old-old participants' average threshold for \baricon with 24 data points was 1005~ms. With that in mind, we lean toward recommending \donuticon for older adults if there are several data points or if comparison speed is critical.
Notably, older adults were consistently slower with \radialicon across all data sizes, with times ranging from 943~ms (7 data points) to 5460~ms (24 data points). This suggests that \radialicon are less effective than Donut and \baricon for quick data comparisons on smartwatches. 

\noindent\textbf{\radialicon is Preferable for Displaying Task Progress and Completion.} Our combined analysis of performance time and participant feedback indicates that \radialicon are possibly more suitable for visualizing progress and task completion (e.\,g., achieving a step goal) for a single data point rather than numerical value comparison. 
While younger adults had average thresholds below 1000~ms with \radialicon for 7 and 12 data points, old-old participants had average thresholds over 1000~ms for all three data sizes. 
This conclusion aligns with Blascheck et al.'s~\cite{blascheck2023studies} work suggesting that \radialicon are better for representing single data points in part-to-whole comparisons, particularly for goal completion.

\noindent\textbf{Balance the Tension between Visualization Size and Data Size.} Our participants frequently noted difficulties in distinguishing graph elements (e.\,g., bars in \baricon) as they became narrower with increasing data size.
Because its region widths are dependent on data values instead of the number of data points, the regions of a Donut chart shrink at a slower, different rate as data size increases compared to the other two chart types. We speculate that older adults' greater reported preference for this chart across data sizes compared to younger adults may partly be due to this phenomenon.
While the tension between display size and information density is a known challenge in visualization~\cite{wu2012visizer}, the small dimensions of smartwatch screens can possibly exacerbate and amplify the issue. 
Hence, visualization designers should ensure adequate perceptibility of information for older adults by either using charts less affected by data size (e.g., Donut charts) or considering recommendations from 
existing work aimed at displaying large amounts of data on small screens~\cite{chen2017visualizing,Neshati2019}.\\\vspace{-1em}

Our current knowledge of empirically derived glanceable visualization design guidelines for older adults is lacking. To address this issue, further work is needed to examine a broader range of glanceable visualizations and analysis tasks.

\section{Limitations and Future Work}
In replicating  Blascheck et al.'s~\cite{blascheck2018glanceable} study, we inherit limitations the authors originally noted. They discuss that some of the aspects of the study design, such as the color variance and the placement of the dot markers, could have influenced participants' performance and contributed to the differences observed between \donuticon, \baricon, and \radialicon. It is possible that these factors similarly impacted the performance of the older participants in our study. 
They also mention the simplicity of the studied comparison task, requiring the participants to choose which of two marked elements displayed a larger value. In contrast, more complex and perceptually demanding tasks could increase observed thresholds. Similarly, the performance thresholds of our older participants are likely to change with more complex tasks. 
Moreover, reading the watch on a stand at an ideal height while sitting is not representative of real-world use, as worsening hand-eye coordination due to aging could introduce difficulty in reading a worn smartwatch~\cite{guan2000effect}. Further work can ascertain the impacts of worsening hand-eye coordination on reading worn glanceable visualizations. 

Our study's participant pool was notably skewed toward higher educational levels, predominantly comprised of individuals with master's degrees. This almost certainly arises from our recruitment strategy, which focused on local participants in an area rich in educational institutions; 10 of our 24 participants stated a primary or most recent educational occupation. This skew, however, aligns with the original study, which also had a master's degree as the most common educational level (see \autoref{tbl:educ-lev-comp} for more details). While the higher academic levels could correlate with higher visualization literacy, this demographic may more accurately reflect the likely user base of smartwatches, as age and educational level are influential factors in technology adoption~\cite{rupp2018role}. 

The context, in which a conceptual replication study occurs (e.\,g., different time, place, or with other participants) can offer both benefits and challenges. On the positive side, these variations can enhance the generalizability of the findings and provide robust tests of underlying theories across different conditions, thereby enriching scientific understanding. However, such changes may also introduce potential confounding variables. In \autoref{sec:aging-agility}, we reflect on the observed difference between younger and older adults' performance from the perspective of age-related perceptual and cognitive disparities between the two groups. However, other differences between younger and older adults could have also impacted the results. For instance, life experience may notably vary between these age groups, influencing their reactions to experimental stimuli or tasks. Social and cultural norms also differ between generations, potentially affecting behavior and attitudes in ways that complicate the study's interpretation. 

The primary focus of this work was to establish and compare older adults' time thresholds. There is still an unmet need for more extensive investigation of various aspects of glanceable visualization design for this population. For instance, more work is needed to understand the design considerations for presenting multiple visualizations on a watch face (recently done with younger adults by Blascheck et al.~\cite{blascheck2023studies}) as well as how to best present text and fine-grained visual components that may be difficult to see with low visual acuity on such a small screen~\cite{mitzner2015considering}. 
Methods of annotating data and bringing attention to noteworthy data points likely require different considerations on a smaller screen and for shorter viewing times (glances), with previous work recommending these types of visual aids for older adults~\cite{le2015evaluation}. Some participants in our study mentioned a preference for viewing data as numbers instead of visualization on their watch. This may indicate that more precise low-level tasks such as \textit{retrieve value} and \textit{determine range} may better represent numbers than a complete visualization. Facilitating rapid interactions for glanceable visualizations is a noted challenge within this area~\cite{blascheck2021characterizing}. Investigating which low-level tasks are even suitable for glanceable visualization is also an open problem that could help narrow the research community's focus.

\section{Conclusion}
In this paper, we replicated the study by Blascheck et al.~\cite{blascheck2018glanceable} that investigated how quickly people can compare data using a glanceable visualization on smartwatches. Our primary objective was to rerun this study with older adults (age~$\geq 65$) to establish time thresholds and compare the results of the two studies (original and current) to learn about possible differences between the younger and older adults' performance, strategies employed to perform tasks, and if any differences existed concerning chart type. For each combination of chart type (\baricon, \donuticon, \radialicon) and data size (7, 12, 24), participants ($n=24$) completed 9 staircases while performing a data comparison task using a two-alternative forced choice approach. Our results showed weak or strong evidence of differences between younger adults (original study) and old-old adults (age $\geq 75$) across almost all conditions. However, all but one condition (\radialicon~24) had insufficient evidence of differences for young adults and young-old adults (age 65-74). 

These results prompt interesting questions regarding how we study glanceable visualization for older adults, especially when defining who an ``older adult'' is, as well as whether differences in performance speed have noticeable real-world consequences. We also discuss our experience working with older adults in an in-person study, including takeaways regarding flexible study design and a brief reflection on working with vulnerable populations. Future work can expand on these questions and contribute further understanding of how we can better design glanceable visualizations for older adults. We hope this study sparks additional interest in the visualization community for considering equitable design in areas that, while more challenging, can level the playing field of data-driven insights.

\begin{acks}
Tanja Blascheck is funded by the European Social Fund and the Ministry of Science, Research and Arts Baden-W\"urttemberg.
\end{acks}

\bibliographystyle{ACM-Reference-Format}
\bibliography{bibliography}

\end{document}